\documentclass[journal]{IEEEtran}
\usepackage{amsmath,amsfonts}
\usepackage{algorithmic}
\usepackage{algorithm}
\usepackage{array}
\usepackage[caption=false,font=normalsize,labelfont=sf,textfont=sf]{subfig}
\usepackage{textcomp}
\usepackage{stfloats}
\usepackage{url}
\usepackage{verbatim}
\usepackage{graphicx}
\usepackage{cite}
\usepackage{cleveref}
\usepackage[tight-spacing = true, separate-uncertainty=true,multi-part-units=repeat,range-phrase = --,exponent-product=\cdot, range-units = repeat,range-phrase=\,\,to\,\, ]{siunitx}
\usepackage{newunicodechar}
\usepackage{units}
\usepackage{color, soul}
\usepackage{float}
\usepackage{bm,bbm,dsfont}
\usepackage{calrsfs}
\usepackage{empheq}
\usepackage{esdiff}

\usepackage{footnote}
\usepackage{breqn}

\usepackage{multirow}
\usepackage[switch, modulo]{lineno}
\expandafter\let\csname equation*\endcsname\relax
\expandafter\let\csname endequation*\endcsname\relax
\usepackage[font=footnotesize]{caption}
\usepackage[separate-uncertainty=true,multi-part-units=repeat,range-phrase = --,exponent-product=\cdot, range-units = repeat,range-phrase=\,\,to\,\, ]{siunitx}
\newcommand*{\smallrel}[2][.8]{%
\mathrel{\mathpalette{\smallrel@{#1}}{#2}}%
}

\usepackage[colorinlistoftodos]{todonotes}

\begin{document}
\setlength{\abovedisplayskip}{6pt}
\setlength{\belowdisplayskip}{6pt}

\title{Deploying the high-power pulsed lasers in precision force metrology -- SI traceable and practical force quantization by photon momentum}
\author{Suren Vasilyan, Thomas Fr\"{o}hlich, Norbert Rogge, 

Institute of Process Measurement and Sensor Technology, Technische Universit\"at Ilmenau, 98683 Ilmenau, Germany 

Corresponding author's email: suren.vasilyan@tu-ilmenau.de

}


\maketitle
\begin{abstract}
Design and operational performance of table-top measurement apparatus is presented towards direct Planck constant traceable high accuracy and high precision small forces and optical power measurements within SI unit system. An electromagnetic force compensation weighing balances, highly reflective mirrors and high-energy pulsed laser unit (static average power \SI{20}{\watt}) are tailored together with a specially developed opto-electro-mechanical measurement infrastructure for cross-mapping the scale-systems of precision small force measurements obtained with a state-of-the-art classical kinematic system employing Kibble balance principle in the range of \SIrange{10}{4000}{\nano\newton} in comparison with forces generated due to quantum-mechanical effect namely the transfer of the momentum of the photons from a macroscopic object. Detailed overview of the adapted measurement methodology, the static and the limits of dynamic measurement, the metrological traceability routes of the measurement parameters, quantities and their measurement uncertainties, parametric up(down)-scaling perspectives of the measurements are presented with respect to the state-of-the-art measurement principles and standard procedures within the newly redefined International System of Units (SI).
\end{abstract}


\section{Introduction}
\begin{figure}[ht!]
\centering
\includegraphics[width=0.95\linewidth]{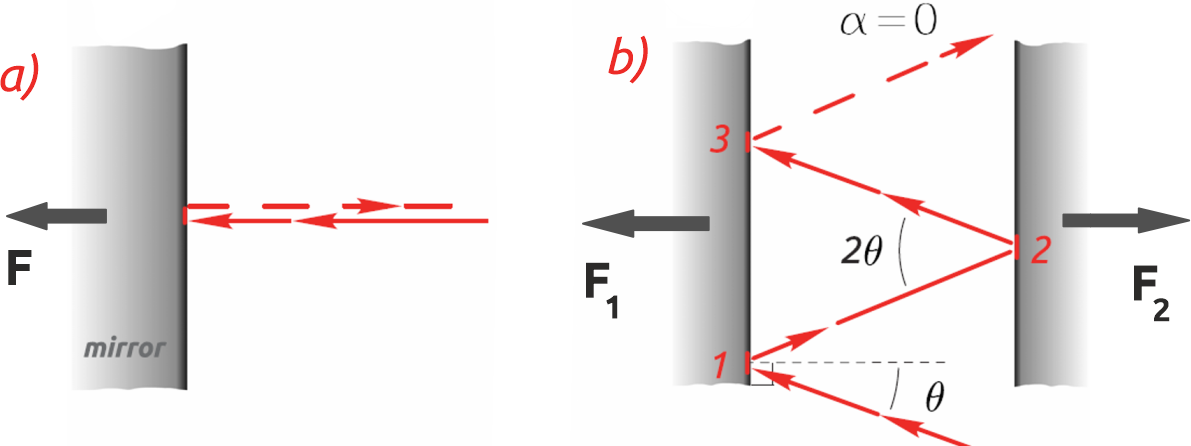}
\caption{2-D geometry of the laser beam and flat surface mirrors. a) single reflection where the incidence angle of the laser beam is parallel to the normal of the surface, b) multi-reflection within two quasi-parallel mirrors with relative $\alpha$ angle between the mirrors and the incidence angle of the laser beam denoted by $\theta$.}
\label{fig:Fig1_mirrors_reflection}
\end{figure}

\IEEEPARstart{T}{he} redefinition of the International System of Units (SI) stimulated a new quest for developing ever improved methods supporting the calibration tasks in the fields of precision mass/force metrology and 'unsurprisingly' in high power optical (laser energy) metrology \cite{Borys_PTB, knopf2019quantum, Williams_2019, Yuan_2020, Pinot_french_overview, Rothleitner_German_pamphlet}. Specifically, as of the recent advances in macroscopic realization of the quantum-based kilogram, defined by a natural constant and emerged from quantum mechanics -- namely the Planck constant which in itself is already numerically exactly defined as $h$=\SI{6.62607015e-34}{\joule\per\hertz}, -- guiding methods arose for improving the accuracy of quantification of the optical power of high energy lasers covering the range from \SIrange{e0}{e5}{\watt} and above, and for generation of reference calibration small forces ranging from \SIrange{e-5}{e-8}{\newton} and below \cite{PB2_tm, Vasilyan_2021, Williams_2, Keck_Schlamminger_100kW}. One of the methods reintroduced in the last decade is the quantum-based realization of small forces using the effect of the momentum transfer of the absorbed and re-emitted photons from the ultra-high reflective mirrors -- the radiation pressure \cite{Nichols_Hull}. These forces are small, and scale as approximately \SI{6.7e-9}{\newton} for the unit optical power of the laser field generated upon its single reflection at normal incidence from a perfectly reflecting material. The use of the photon momentum together with the modern state-of-the-art conventional and quantum measurement techniques may serve as a unifying platform for combining the mass/force- and optical metrology fields. The underlying fundamental physical principle is the conservation of the momentum/energy, and in practice, the realization is reduced to precisely measuring the instantaneously generated force which is proportional to applied optical power, as a result of total momentum transfer of the photons when a single \cite{Williams_2} or multiple reflections \cite{OPEX_Vasilyan} of the laser beam from the highly reflective mirror(s) occur (see \cref{fig:Fig1_mirrors_reflection}). The multiple reflection case introduces an enhanced, amplification effect of generated forces due to the occurrence of isochronous reflection events. Note that for a single source of electromagnetic radiation (laser) the multi-reflections occur sequentially and with the speed of the light for both coherent and incoherent (i.e. non-overlapping or specular) reflections \cite{OPEX_Vasilyan,Manske,Vasilyan_2021}. 

In conventional optical metrology the optical power of lasers at the \SI{}{\watt}, \SI{}{\milli\watt}, \SI{}{\micro\watt} levels can be measured very accurately with thermopile- or silicon (Si) photodiode reference detectors (with the avalanche photodiode detectors the laser intensity can be calibrated against electrical power and thus against Planck constant statistically even at the single-photon level), therefore, the force measurements at the \SI{}{\nano\newton}, \SI{}{\pico\newton}, and \SI{}{\femto\newton} levels can be referenced very accurately with the lowest measurement uncertainties. By virtue of the same relation, the photon momentum-based standardized force measurements can be used to develop a viable, accurate, and absolute optical power meter at higher ranges, on the order of \SI{}{\watt}'s to several hundred \SI{}{\kilo\watt} and above, with direct SI traceability. The formerly presented reports show that the photon-momentum method allows the quantification of the magnitudes of both the small force and high optical power/energy with superior accuracy and precision in comparison with other methods that are practiced currently \cite{Vasilyan_2021}. It is also evident that it paves a road towards opportunities for broadly practicable, frequency-resolved, simplified SI-traceable, and truly quantized force sensing as quantified by the Planck constant. And, resolving similarly for the optical power measurements for continuous-wave (CW) or high-pulse-energy high-repetition-rate lasers \cite{Manske}. 

\section{Outline of paper}

In this contribution, we present a multi-functional table-top metrological apparatus with a goal to introduce systematically the core concept of the photon-momentum-based SI-traceable measurement infrastructure for generation and measurement of small calibration forces below \SI{10}{\micro\newton} range, referenced by the pre-characterized opto-electro-mechanical system and applied optical power of the laser field (referenced by the conventional optical detector) at several ten W power level. Particularly, we will be making use for the first time of the pulsed laser sources and the vast optionalities of the same in generating any desired type of the input function optically. This concept and the measurement infrastructure can be adapted for its second main purpose, an inverted measurement routine, namely as an optical power detector for high power lasers at \SI{1}{\watt} up to several \SI{100}{\kilo\watt} levels provided the force member(sensors) are calibrated and well-characterized for the small force measurements at \SI{10}{\micro\newton} to \SI{}{\nano\newton} level and below.

The table-top metrological system is designed and constructed using two identical state-of-the-art weighing balances working on the principle of electromagnetic force compensation (EMFC) method for precision mass measurement. Each of the two EMFC balances appears in oversimplified form as a simple pendulum operating in its naturally achievable stable equilibrium state for detection of horizontally acting forces. Under the construct of the differential signal, by combining the operation of both EMFC balances together, a reduction of common-mode noise due to tilt or accelerations or temperature changes is obtained. They perform a common force calibration measurements routine based on Kibble balance principle \cite{Kibble, PB2_tm} supported by the necessary opto-electrical measurement infrastructure to enable the SI traceability of the main measurement quantities. The low stiffness of the flexure-based parallel-beam guiding mechanism, the high lever ratio of force transmission, and the extensively well-characterized coil and closed-circuit magnet system assembly nested inside the EMFC balances provide extremely high force sensitivity and reproducibility. Highly reflective mirrors are rigidly fixed  in a vertical plane from both EMFC balance pans and face to each other with their reflective coated sides. This mechanical arrangement allows to prevent the double pendulum effect and exclude the force due to gravity acting on mirrors or mass pan from the general measurement equation. By such geometrical constraint of the macroscopic optical-cavity the specular type of multi-reflection configuration of the laser beam is achieved. Two different pulsed laser sources based on gain current regulated (against the external source developed in-house or internal source developed by OEM's) diode-pumped solid-state technology are used at \SI{532}{\nano\meter} wavelength with approximately \SI{8}{\watt} and \SI{20}{\watt} maximum output averaged optical power at their optimal operational level. The pulse-repetition rate range from \SI{10}{\kilo\hertz} to \SI{650}{\kilo\hertz}  and the pulse width range from \SI{1}{\nano\second} to several \SI{}{\micro\second}. Additionally, we couple these laser systems in the external control mode to an external signal generator in order to accurately control the parameters of the output laser signal, i.e. modulation of the frequency and pulse width. The function generator used in Kibble balance measurement to trigger in burst mode the synchronous voltage measurement routine (including the voltage of optical position sensors or the interferometric signal) serve us in a similar manner to modulate an output optical power signal. In this configuration, we program electronically the laser system to deliver desired frequencies of the pulse trains at specific burst parameters (e.g., rate, duration, etc.), including the control over peak power (amplitude) and the pulse width, latter by adjusting the duty cycle of the signal. Similarly, the frequency counter is used to directly monitor the accuracy of the actual modulated output burst signals from the function generator for the frequency rate ranging at \SI{1}{\kilo\hertz} to \SI{1}{\mega\hertz}. Both, the frequency counter and function generator are locked to an external source of frequency standard at \SI{10}{\mega\hertz}, using a GPS disciplined oven stabilized quartz oscillator with \num{e-10} temporal stability per second (typically better than \num{<e-8} over a time period of hours).

In pursuit of systematics, the above-described core technicalities of the table-top system will be broken down into the following blocks. We begin by presenting in \cref{theo_back} the theoretical background underlying the concept of photon-momentum-based force and optical power measurements considering the case of specular reflection. Then we revisit the basics of the operational principle and defining equations for the realization of the quantum-electrical-based measurement routine via Kibble balance principle in \cref{PB2_and_merging} which is adapted to the developed setup, and how this routine is theoretically extended to operate together with high power laser units. In the \cref{note_overview} we briefly describe the most notable state-of-the-art achievements in this field of metrology and several crucial real-practical limits and design-specific considerations. Based on the rationale provided in the former section the design of the photon-momentum-based Planck's-constant-traceable force measurement setup will be introduced in \cref{design}. Further, we will detail on progress in the developed measurement infrastructure, the systematics of the designed measurements, and a few results from a set of extensive measurements. We conclude the paper by discussing the improvements towards the development of a robust SI-traceable measurement routine in order to enable a more detailed systematic evaluation of the measurement quantities in the future. Note that few aspects with respect to the future improvements, the full characterization of the influences that are contributing to measurement results, the options to correct or diminish such influences, identification of other errors sources, and generally, a comprehensive statistical and systematic evaluation of the measurement data, not accounted in this paper are subject to work in progress.

\section{Theoretical background}
\label{theo_back}
The photon momentum of light as a fundamental quantity can be described in simple terms by the energy of the incident photons (i.e. flux or stream) and be defined by Planck's constant $h$ and the wavelength/frequency of light over the time of interaction. With the current state-of-the-art quantum-based technologies, it is already possible to conduct for the systems at few photon level an accurate frequency and time measurements with a relative measurement uncertainty of \num{e-16} and below, yet more practically at the industry-applicable level of realization the measurement uncertainty is commonly achieved at the range of \num{e-11} to \num{e-14}. For very low optical powers, by using the methods from quantum optics and basic photometry it is possible to measure/count the photons statistically at a rate of approximately \num{1000} photons per second. Due to this reason, the single-photon avalanche diode technology gradually becomes the industry standard and is ideal for extreme low-level light detection \cite{Kuck, Georgieva_Kuck_Lopez}. The common commercial solutions provide somewhat a rate of up to \num{10000} photons per pulse (approximating to \SI{}{\femto\watt} power level) for lasers having about \SI{<1}{\nano\second} to \SI{50}{\nano\second} pulse width. Therefore, they are widely employed in optical metrology as optical power detectors, in the optical communication industry, specifically for Quantum Key Distribution technologies, and in other quantum sensing systems. Furthermore, by the use of state-of-the-art cryogenic primary standards as low as \SI{}{\nano\watt} orders of powers can be detected, whereas the upper limit is typically given for below \SI{1}{\milli\watt} power level with an expanded measurement uncertainty of about \SI{0.002}{\percent}. Despite the high accuracy in measuring the time and frequency at the single-photon level and quantifying low flux rates, these approaches are not viable at the higher optical flux rates and to quantify the number of photons statistically becomes an impractical task, e.g. as in \SI{10}{\watt} laser power case for \SI{532}{\nano\meter} wavelength it scales at the level of $>\num{e19}$ number-of-photons (Nph). Therefore, the optical power measurements at those high power levels and above typically are complemented by the use of technologically other classes of detectors (conventional Si-diode and thermal) that are limited already to the percentage level of total expanded uncertainty. An approach to measuring the effective force \textit{\textbf{F}} produced by the momentum of photons, $\textbf{p}$ \cref{eq:2,eq:1}, of the laser field, offers a partial solution to this challenge in order to reduce the uncertainty of the power measurements at those high power levels. Additionally, the method allows to directly co-relate the mechanical and optical excitation frequencies at the fundamental physical level thereby extending the frontiers of other practical measurement aspects applied elsewhere, e.g. in force metrology. 

\begin{figure*}
\centering
\includegraphics[width=0.9\linewidth]{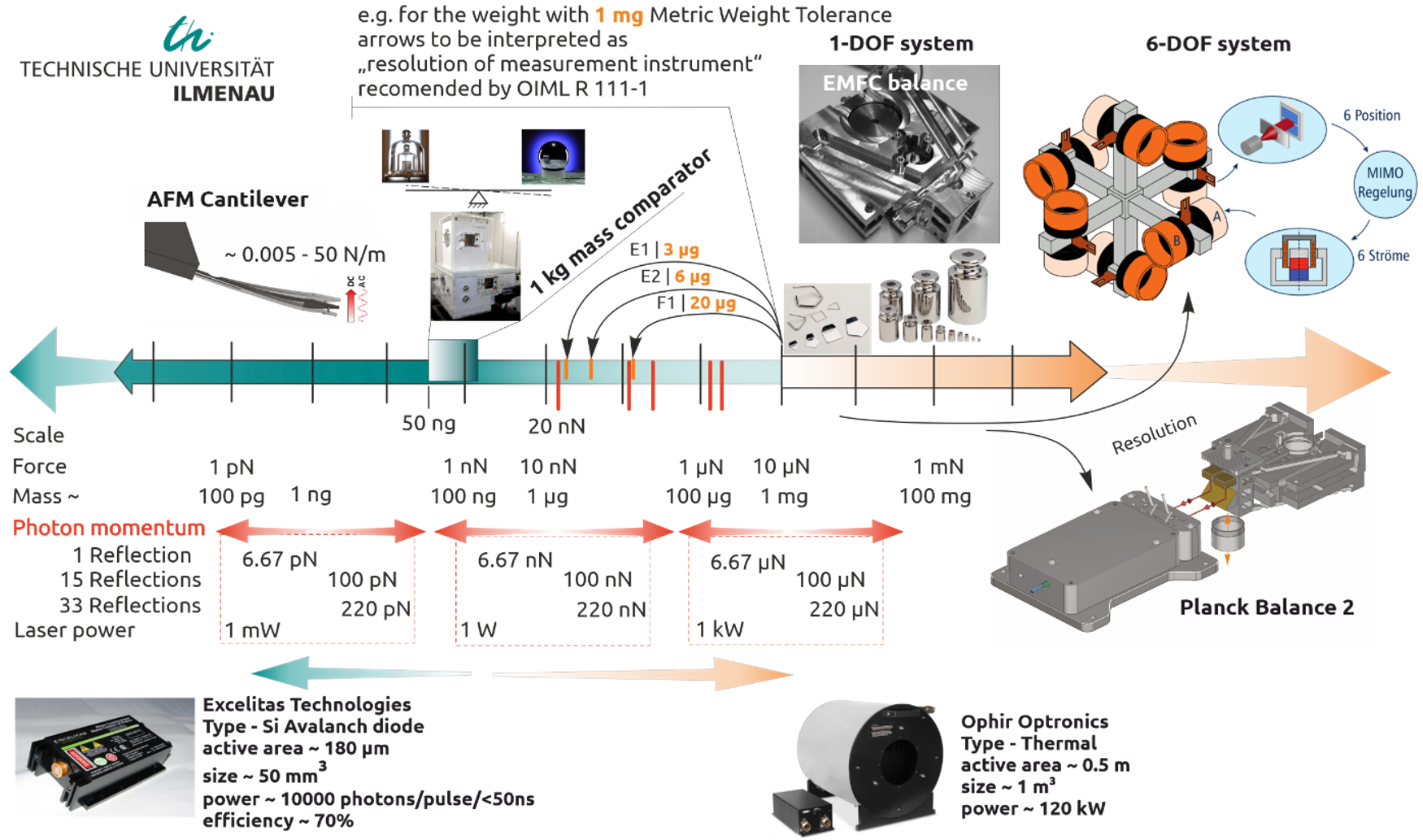}
\caption{The sketch represents the technological map and the scale overlap for mass, force, and optical power measurements when employing the photon momentum method. At the upper panel are shown some of the commonly used state-of-the-art apparatuses and techniques for mass and force measurements, including the \SI{1}{\kilo\gram} mass comparator \cite{Mass_comparator} by which about \SI{50}{\nano\gram} (\SI{50}{pN}) standard uncertainty is achievable for \SI{1}{\kilo\gram} mass comparisons. At the lower part, the optical power detectors for the single-photons using the avalanche photodiodes and for powers about \SI{100}{\kilo\watt} using thermal power meters.}
\label{fig:Fig0_tech_map}
\end{figure*}
 
\begin{equation} 
\textbf{F} = \diff{\textbf{p}}{t} =\frac{\mathsf{E'}}{c}   \diff{\textbf{n}}{t} = \frac{\mathsf{E'}}{c}  \frac{\Delta\textbf{n}}{\Delta t}=  \frac{\text{Power}\,'}{c} \Delta \textbf{n}
\label{eq:1}
\end{equation}

Here, ${\mathsf{E'}}$ is the total energy of a single photon, whereas its radiant energy/flux per unit of time defines its optical power $\text{Power}\,' = {\mathsf{E}'/\Delta t} $, c is the speed of the light, \textbf{n} is a unit vector indicating the direction of the photon’s motion. For the given optical power of the laser $\text{Power} = \text{Nph} \cdot \text{Power}\,'$ and in the case when the angle of incidence of the laser beam is parallel (the angle of incidence is $\theta\approx0,\ \cos{\theta}=$1) to the normal of the surface of an ultra-reflective mirror, $R_L\rightarrow1$ (negligible transmission and absorption), the measurement equation is reduced to the following relation
\begin{equation}
\boxed{F= \frac{\text{Power}}{c} (1+R_{L})}
\label{eq:2}
\end{equation}
Here, the power could be interpreted as the constant optical power for the CW laser over a certain time scale or the average power of pulsed laser energy obtained by evaluating the integral of the pulse function over a defined time scale. For example, taking the interactions at the level of macroscopic objects for a high reflective mirror (assuming an ideal reflector $R_L$  \SI{>99.999}{\percent} defined for the specific wavelength value), the force exerted on a mirror by a CW laser source with \SI{15}{\milli\watt} average optical power can be approximated theoretically as \SI{100}{\pico\newton}, which is approximately equivalent to the gravitational force acting on the 0.01-µg-mass piece, determined as
\begin{equation}
F= m g
\label{eq:3}
\end{equation}
Due to technical limitations, the manufacturing reliability and the handling of such small calibration masses, this kind of direct comparison is practicable only for the laser powers at the kW level with low resolution. Note the mass measurements/calibration are routinely performed at the level of \SI{1}{\milli\gram} (\SI{10}{\micro\newton}) and above. For the mass values below this level relevant for comparison proposes with photon-momentum forces the measurements and their calibration are cumbersome. To overcome this obstacle, a method with a multi-reflected laser beam configuration has been introduced to amplify the total net force generated by the same amount of laser power \cite{Vasilyan_2021,OPEX_Vasilyan,Manske}, which have been further adapted in \cite{Shaw_2019, Artusio_Glimpse_2021}. The total force represented by the sum of contributions after each reflection event occurs is 
 \begin{multline}
F_{T}=\sum_{i=1}^{N}F_i=\frac{(1+R_L)}{c}\sum_{i=1}^{N}{\rm Power}_i, \\
{\rm with}\ \ \ \ \ \sum_{i=1}^{N}{\rm Power}_i={\rm Power}_1\sum_{i=1}^{N}{R_L}^{i-1}
\label{eq:4}
\end{multline}
where $i=1,2,\ldots,N$ is an integer value showing the number of specular reflections (see \cref{fig:Fig1_mirrors_reflection}), ${\rm Power}_1$ and ${\rm Power}_i$ are the laser power at the input (first reflection, hence ${\rm Power}_1={\rm Power}={\rm Power}_{input}$) and at the $i$th reflection event (at $i=N+1$, ${\rm Power}_{N+1}={\rm Power}_{output}$). The second part in \cref{eq:4} represents the magnitude of the total optical power recycling in the cavity or the sum of residual optical powers after each reflection event occurs. In accordance with this theoretical calculation, the force produced by a 15-W-average-power laser source with 20 reflections is approximately \SI{2}{\micro\newton} (\SI{0.2}{\milli\gram}) for surfaces with \SI{99.999}{\percent} reflectivity, and to \SI{10}{\micro\newton} (\SI{1}{\milli\gram}) when the power is \SI{75}{\watt}, whereas for \SI{15}{\milli\watt} and \SI{75}{\milli\watt} powers with a single reflection, it produces only  \SI{100}{\pico\newton} (\SI{0.01}{\micro\gram}) and  \SI{500}{\pico\newton} (\SI{0.05}{\micro\gram}) forces, respectively. In practice, we have already demonstrated \cite{OPEX_Vasilyan} that it is possible to increase the magnitude of these forces by several orders by using a special multi-reflected laser beam configuration within the passive/active cavities. This makes the photon momentum method ideal for generating reference force values at the higher orders of magnitude for the verification purposes with respect to other well-established and conventionally accepted gravimetric method. In \cref{fig:Fig0_tech_map} the red marks on the scale highlight the values of the force realization at approximately \SI{20}{\nano\newton}, \SI{150}{\nano\newton} \cite{OPEX_Vasilyan}, \SI{350}{\nano\newton} \cite{Manske}, \SI{2000}{\nano\newton} \cite{Vasilyan_2021}, and \SI{4000}{\nano\newton} (current work) level as milestones obtained by the photon momentum method at the PMS Institute.

The \cref{eq:2,eq:3,eq:4} show that the optical power measurement, specifically at \SI{1}{\watt} level and above, can be made with a simple and short traceability chain directly connecting it to the base/derived units, and to the mass standard within the early version of the SI unit system. However, knowing that
 \begin{itemize}
\item mass realization is defined by the numerical value of the Planck’s constant $h$, and
\item it is necessary to measure a force instead of the mass,
 \end{itemize}
then, the use of the calibrated mass pieces can be avoided entirely. Thus, one may consider connecting the optical power measurements directly to Planck’s constant through the force measurements, with the use of a concept specially developed for calibrating mass pieces, the Kibble balance (KB) principle \cite{Kibble}. Effectively, via a reduction of one intermediate step and by replacing the “\textbf{force mode}” of the KB principle with the “\textbf{photon momentum mode}” or “\textbf{momentum mode}” \cite{Vasilyan_2021} instead, the value of the mass, and therefore its measurement uncertainty, will not enter the general measurement equation. Owing the progress achieved in recent years in developments of tabletop versions of the KBs, e.g. Planck-Balance 2 (PB2) \cite{PB2_tm, Froehlich2020neue} or KIBB-g1 \cite{chao2020performance}, the realization of the optical power (or small calibration force) measurements can be undertaken with similar systems, where the requirements on the level of the uncertainties attached to each measurement quantity are relaxed in comparison with KB systems. As an example, with the PB2 apparatus, typical measurement uncertainties required for industrial mass calibration in accordance to \text{OIML R 111-1} for weights of E2 class \cite{OIML} is achieved. This allows measuring mass artifacts in the \SI{1}{\milli\gram} to \SI{100}{g} range in the laboratory without access to any calibrated masses, instead with direct traceability to the Planck constant by quantum electrical measurements. Note that the calibrations are typically performed for a specific set of nominal masses values, however, with the PB2 system \cite{PB2_tm} we demonstrated also by 'quasi-blind' measurement campaign using a set of uncalibrated mass pieces that it is capable of mass calibrations for other mass values along the continuous scale. Additionally, it is delivering the necessary level of measurement uncertainty required by universally accepted guidelines \text{OIML R 111-1} \cite{OIML}. The advantage of implementing measurements without the “force mode” is that the knowledge of the value of the gravitational acceleration is not required at all, meaning that the measurements can be done independently of the gravitational field at the site of the measurement and not necessarily in the vertical motion direction, but for the other geometrical arrangements (horizontally) as well.

\section{Operational principle and defining equations}
\label{PB2_and_merging}

The basis for the development of an SI-traceable framework for the measurements of mass, force, and the optical power of the lasers via photon momentum-based force generation and quantum electrical measurement is a special measurement routine that uses the KB principle \cite{Kibble}. This principle, in recent decades promoted for determination of Planck’s constant and further adapted for determining the mass and dissemination of the quantum-based kilogram, has been extensively studied since the first reports in the late 1970s [7]. The KB (formerly known as Watt balance) relies on precise virtual comparison of mechanical and electrical powers obtained from two-step experiments both based on electromagnetic interaction and with the use of macroscopic quantum effects, the Josephson and the quantum Hall effect. 
\begin{figure}[ht!]
\centering
\includegraphics[width=0.7\linewidth]{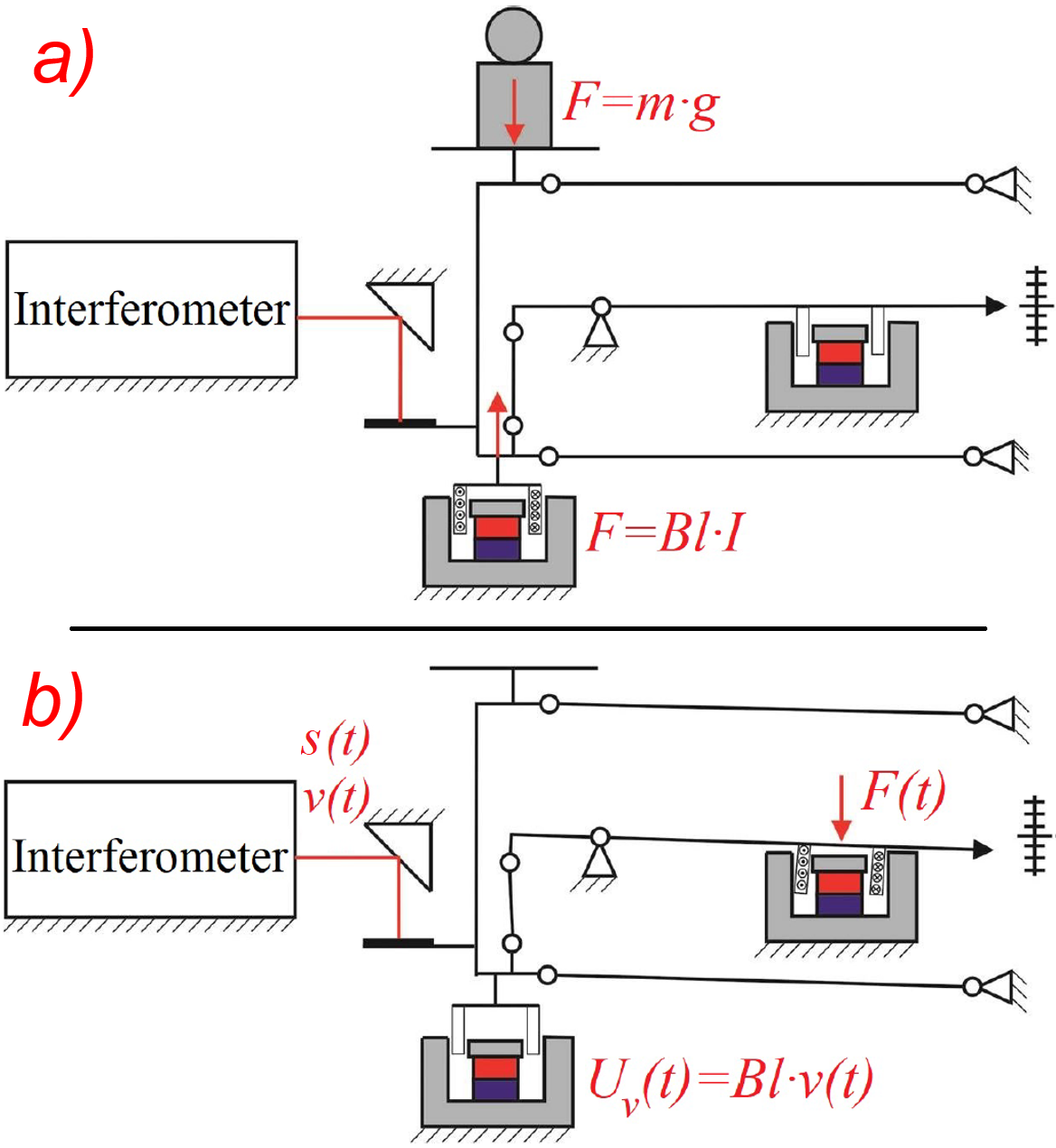}
\caption{Schematics of Kibble balance principle integrated in the PB2 setup, a) force mode, b) velocity mode. The principle is adapted for the table-top setup presented in current work for measurements of forces in horizontal direction.}
\label{fig:Fig2_planck_balance_2}
\end{figure}
For such comparison, a two-step sequential measurement is carried out by measuring electrical quantities in order to determine the electromagnetic force factor of the magnet and the coil system, that is integrated in the core of the experimental setup. These measurements, named force/static mode and velocity/dynamic (\cref{fig:Fig2_planck_balance_2}) mode, are described by
\begin{equation}
    F=F_f=F_g=m \cdot g=Bl \cdot I=Bl\cdot\frac{U_f}{R},
    \label{eq:Fmg}
\end{equation}
and
\begin{equation}
    U_v=Bl\cdot v\ .
    \label{eq:Uv}
\end{equation}
The gravitational force $F_g$ (\cref{eq:3}), acting along the gravitational acceleration $g$ direction on the mass $m$ to be determined, is counterbalanced by the electromagnetic force $F_f$ generated by an electric current $I$ applied to a coil with length $l$ situated in the magnetic field $B$. From the force mode (\cref{fig:Fig2_planck_balance_2}a), the current can be measured as the voltage drop $U_f$ over a precision resistor $R$. For the velocity mode (\cref{fig:Fig2_planck_balance_2}c), the induced voltage $U_v$ is measured when the same coil is moved in the same magnetic field with velocity $v$. In practice, the link between the measured mass and Planck’s constant can be established by calibrating the voltmeters and the precision resistor against the programmable Josephson voltage standard (PJVS) and the quantum Hall resistor (QHR) when measuring the induced voltage and the voltage drop over a precision resistor, as in the case of Planck-Balance 2. According to the \cref{eq:Fmg,eq:Uv} it follows that this measurement principle can serve also to establish the standard of the force and realize the Newton directly traceable to the Planck constant while omitting the need to obtain the value of the local gravitational acceleration at the site of the measurements \cite{Borys_PTB, knopf2019quantum, Kibble, Froehlich2020neue, PB2_tm, tm_Falko}. Moreover, by replacing the product $m\cdot g$ in \cref{eq:Fmg} by the \cref{eq:2} we obtain the photon-momentum-based force-sensing method, from which the above two \cref{eq:Fmg,eq:Uv} can be solved for the optical power (also reflectance value) to obtain the unifying calibration formula.
In accordance with the operational terminology of the KB, as mentioned also for PB2, the photon momentum-based force measurements (or average optical power of the lasers) can also use two measurement modes —- velocity mode (A) and \textbf{"photon momentum mode"} (B) -— in a periodic manner, creating an ABBA or ABA measurement cycle  similar to common weighing procedures known from metrological guidelines for canceling out the effects of the linear drifts of measurement devices \cite{OIML,Mass_comparator,GUM, DIN}. In such a measurement scheme, the time required for mass/force/optical power calibration can be markedly reduced. As an example, in operating with PB2, the typical measurement time for the single-mode can be adjusted to be approximately at the order of \SIrange{30}{100}{\second}, whereas for achieving even lower measurement uncertainties, the number of repetition of ABBA or ABA cycles can be increased.

Contrary to the common approach that most of the known KB systems use, in our case, the velocity mode measurements are made by moving the coil sinusoidally and measuring an AC \cite{PB2_tm,tm_Falko} rather than DC signal generated in this mode \cite{Kibble,chao2020performance}. The coil oscillates in the magnetic field of the permanent magnets with a frequency ranging approximately \SIrange{0}{10}{\hertz} with an amplitude that can be adapted in the range from several \SI{}{\micro\meter} to \SI{50}{\micro\meter}. The velocity $v$ of the coil motion is to be measured in terms of position/length measurement using an interferometer and reference clock, the induced voltage in the coil $U_v$ using a voltmeter. Note that for very low-frequency cases below \SI{0.1}{\hertz} the measurement process is reduced to DC measurement mode, where the induced voltages in our case would become very small.

Under the assumption that both the position and the voltage measurements are ideal single-component sine waves with amplitudes $S$ and $U_{ind}$ and initial phases $\varphi_s$ and $\varphi_u$, respectively, the oscillation can be described by
\begin{equation}
s\left(t\right)=S_0+S\sin{\left(\omega t+\varphi_s\right)\ },
\label{eq:7}
\end{equation}
where $S_0$ is the DC offset and $\omega=2\pi f_{sig}$ is the angular frequency with $f_{sig}$ denoting the oscillation frequency. The coil velocity therefore can be obtained as the derivative of \cref{eq:7} with respect to the time $t$ as
\begin{equation}
v\left(t\right)=\omega S\cos{\left(\omega t+\varphi_s\right)},
    \label{eq:8}
\end{equation}
and the induced voltage as
\begin{equation}
    u\left(t\right)=U_{ind}\cos{\left(\omega t+\varphi_u\right)}.
    \label{eq:9}
\end{equation}
Assuming that the $Bl$ is constant during the whole range of coil movement it can be computed by dividing the measured amplitude of the induced voltage by the amplitude of the coil velocity as
\begin{equation}
Bl=\frac{U_{ind}}{\omega S}=\frac{U_{ind}}{2\pi f_{sig} \cdot S}.
\label{eq:10}
\end{equation}
The oscillation frequency $f_{sig}$ can be accurately measured by a frequency counter, therefore, the accuracy of identifying the $Bl$ immensely depends on the estimation of amplitudes $U_{ind}$ and $S$.
Thus, by obtaining the $Bl$ factor we may plug the result in the \cref{eq:Fmg,eq:Uv} in order to obtain the force values
\begin{equation}
    \boxed{F=\frac{U_f}{R}\frac{U_{v}}{v}=\frac{U_f}{R}\frac{U_{ind}}{2 \pi f_{sig} \cdot S}},
    \label{eq:11}
\end{equation}
with further rearrangements and plugging \cref{eq:2} as the force generated by the photon momentum, the optical power can be obtained by the following equation
\begin{equation}
   \boxed{{\rm Power}_1=\frac{c}{(1+R_L)} \frac{U_f\ }{R\ }\frac{{\ U}_{ind}}{2 \pi f_{sig} \cdot S}},
    \label{eq:12}
\end{equation}
and, for the multi-reflection configuration with $i=\num{1},\num{2},\num{3},..N$ representing the number of reflections, the following equation should be assumed
\begin{equation}
   {\rm Power}_1=\sum_{i=1}^{N}\frac{{\rm Power}_i}{{R_L}^{i-1}}=\frac{c}{(1+R_L)}\sum_{i=1}^{N}{\frac{F_i}{{R_L}^{i-1}}}.
    \label{eq:13}
\end{equation}
Besides the $R_L$, all of the quantities in \cref{eq:12} can be measured directly and SI-traceable using conventional measurement devices at the level of \num{e-6} (ppm) relative measurement uncertainty for the values typical in such systems as for voltage - \SIrange{0.01}{1}{\volt}, resistance - \SIrange{1}{e3}{\ohm}, frequency - \SIrange{0.01}{10}{\hertz}, length - \SI{30}{\micro\meter} to \SI{1}{\milli\meter} (the value of speed of the light in vacuum $c=\,$\SI{299792458}{\meter\per\second} as one of the seven defining constants of the SI is already exactly defined). The $R_L$ at \SI{>99.9}{\percent} level is measured typically using an indirect method at the level of several ppm to \num{100}\,ppm relative measurement uncertainty, by quantifying the optical transition losses of the mirror for a specifically chosen wavelengths. Moreover, for the reflectance measurements, the optical powers below \SI{}{\milli\watt} level is used typically, and the real physical performance of the coating material of the mirrors is not yet able to be fully studied and characterized against the continuous exposure of the pulsed lasers at extremely high energy levels and high repetition rates, including for the CW lasers at several tens of \SI{}{\watt} to \SI{}{\kilo\watt} average power levels. Furthermore, the purposed photon-momentum method when used in combination with systems having sufficiently high Q-factor (or very low stiffness, high sensitivity)  \cite{quantum_gravity, Teufel_quantum} and avalanche diode optical power detectors \cite{Kuck, Georgieva_Kuck_Lopez} can deliver unprecedented force measurement resolution, providing also a real practical quantization.
\begin{equation}
{\rm Power}=\frac{\mathcal{R}\ }{\eta} \cdot h \cdot f_{photon}=\Phi \cdot h \cdot f_{photon}
   \label{eq:14}
\end{equation}
where $\mathcal{R}$ is the average photon count rate of the photons with $f_{photon}$ frequency, $\Phi$ is the photon flux, $\eta$ is the quantum efficiency of the detector used to cross-reference the measurements, and $h$ is Planck’s constant. Here, the fixed numerical value of the hyperfine splitting frequency of the cesium-133 atom, $\Delta \nu( ^{133}Cs)_{hf_s}$, is required to define the frequency and the photon flux and, therefore, with precise knowledge of the numerical value of Planck’s constant, a compatible SI traceability can be established. Above mentioned should be considered in the limiting cases for very small forces when the laser beam is also considered to be a perfectly coherent monochromatic source (continuous stream of photons) with constant intensity.
In this study, some of the crucial fundamental characteristics of the lasers (for both CW and pulsed lasers) used to generate photon-momentum-based small forces are defined in simplified terms, particularly the transverse beam profile is assumed to be constant over the laser pulse and the laser beam with negligible divergence. Also, during the multi-reflection process within the quasi-parallel mirrors the polarization changes and possibly arising effects are neglected\footnote{Due to the initially designed state of the measurements and the experimental convenience at this stage we have neglected a number of physical effects and factors of the laser, such as the shape of the laser beam profile, the value of the beam-quality and its degradation induced by thermo-optic effects (not only inside the laser unit but also during the multi-reflection process) or mode instabilities, the pulse-quality degradation which are induced by nonlinear effects, the optically induced damages, type of $p$- and $s$-polarization or their combination upon superposition  and therefore resulting difference among the individual pulses, and etc. We take advantage of the opportunities offered by manufacturers of the commercial state-of-the-art laser and selected those lasers available at the market which have the best specifications possible.}. The repetition rate is defined as $f_{rep}=\Delta T_{rep}^{-1}$, where $\Delta T_{rep}$ is the pulse repetition time (or period) and the pulse width - the temporal duration of the pulse - is defined as the full width at half maximum (FWHM) by $\tau$. From which the duty cycle is defined as ${\rm Duty\,cycle}\,=\,\tau \cdot f_{rep}={\tau}/{\Delta T_{rep}}$, Thus, after the integration of the specific pulse envelope in square or Gaussian form over the emission time period $t$ for $- \Delta T_{rep}/2$ to $\Delta T_{rep}/2$ limits\footnote{the pulse enveloped is normalized such that its absolute value in square results in the optical power, also as the pulse lasers to be considered mathematically as continuously repeated, periodical signals the integration is restricted to $\pm {T_{rep}}/{2}$ limits}, we arrive at the pulse energy and average optical power relation as
\begin{equation}
{\rm Power}=\frac{E}{\Delta T_{rep}}= E \cdot {f_{rep}}=\,\,\,\,\,\,\,\,\,\,\,\,\,\,\,\,\,\,\,\,
   \label{eq:15}
\end{equation}
\begin{equation}
\,\,\,\,\,\,\,\,\,\,\,\,\,\,\,\,\,\,\,\,=\frac{P_{peak} \cdot \tau}{q} \cdot f_{rep}={{\rm Duty\,\,cycle}\, \cdot \frac{P_{peak}}{q}} \nonumber
\end{equation}
where $P_{peak}$ is the peak power of the laser, $q$ is a coefficient defined from the temporal form of the pulse. For example, in the case of an ideal square signal $q=\num{1}$, for an ideal Gaussian it approximates to $q\approx\num{0.94}$, in general for each particular type of the laser its pulse shape should be found by measuring with possibly high temporal resolution and then be calculated accordingly, or alternatively be solved theoretically assuming its complex pulse envelope.

Similarly, on the practical realization level for the modulated laser pulse, \cref{eq:15}, both the repetition rate $f_{rep}$ and the temporal duration of the pulse $\tau$ (pulse width) can be controlled very accurately using conventional devices, i.e. by a function generator with ultra-high stability timebase up to hundreds of \SI{}{\mega\hertz} level, or measured by frequency counters even at higher precision levels. Moreover, if locked to GPS disciplined high stability clocks or if the laser itself is referenced with the frequency combs, frequency stability at \num{e-10} up to \num{e-13} and above levels can be obtained with negligible drift over the time periods of several hours, similar to the calibration process applied to metrology lasers for length measurement. 
Thus, plugging the \cref{eq:15} into equation \cref{eq:12} and solving for $E$ or $P_{peak}$, the guiding measurement equations for pulse energy or for average peak power can be obtained. 
\begin{figure}
    \centering
    \includegraphics[width=0.4\linewidth]{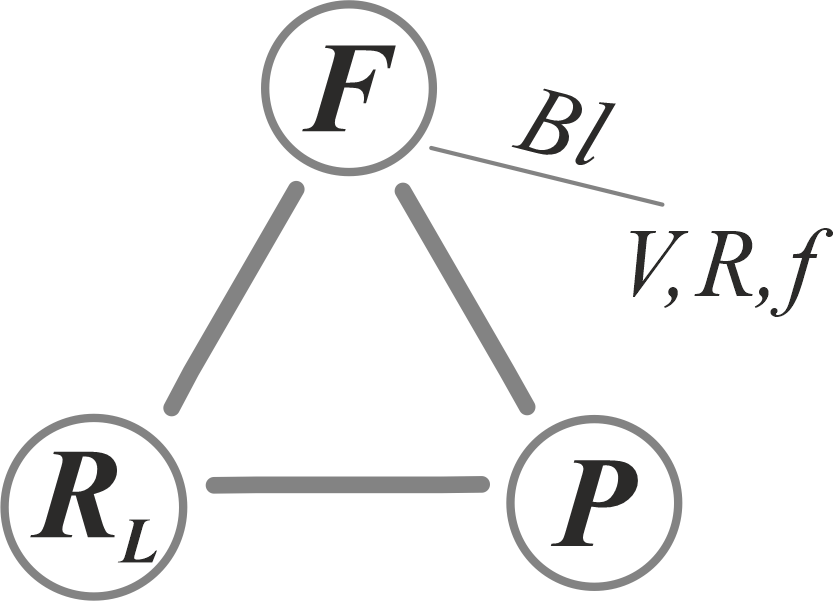}
    \caption{The metrological triangle.}
    \label{fig:triangle}
\end{figure}

Based on theoretical framework discussed above in this section (which virtually can be reduced to schematics given in \cref{fig:triangle}), the following 3 possible measurement/calibration scenarios can be outlined

\begin{itemize}
\item  \textbf{Reflectance} value measurement/calibration (direct and fully SI-traceable)
\begin{equation}
   \boxed{ R_L =\frac{c}{{\rm Power}_1} \frac{U_f\ }{R\ }\frac{{\ U}_{ind}}{2 \pi f_{sig} \cdot S} - 1}
    \label{eq:16}
\end{equation}
\item  \textbf{Optical power} measurement by using \cref{eq:12} for single- or \cref{eq:13} for multi-reflection cases, including the measurements for the pulse energy and peak power.
\item  \textbf{Force} measurement/calibration by using \cref{eq:11,eq:2}. If the average optical power and reflectance value $R_L$ are fixed very accurately by other conventional reference methods and are introduced into the actual measurement process, then both $U_f/R$ and $U_{ind}/2 \pi f_{sig} \cdot S$ quotients in \cref{eq:10} can be generated in a predictable manner and further be measured with highest possible accuracy and precision. Note also that \cref{eq:11,eq:2} provide pulse-width- and repetition-rate-tunable force-sensing optionality while keeping the energy of the laser pulse constant This means that the force values can be referenced fully optically, by changing the average power of a pulsed laser that in turn is referenced by measuring it with conventional power meters continuously, as quasi-CW laser power.
\end{itemize}

\section{Technical overview of notable achievements}
\label{note_overview}

The design of the system was motivated primarily by the necessity to develop a transportable - a table-top - precision force measurement system with robust SI-traceable quantum-electrical measurement infrastructure in order to guarantee the reliability of the measurement data, to extend the operational range of the photon-momentum generated force measurements, and to reduce sources of the errors apparent previously. A number of devices of similar class are known from the literature that are using photon-momentum method adapted for different metrological purposes, having different force and optical power measurement ranges, varying measurement configuration, etc. In \cite{Williams_2} the system arranged in an atypical geometrical operation state for single reflected laser beam configuration demonstrated a relative expanded measurement uncertainty of approximately \SI{1.6}{\percent} for optical power levels between \SI{1}{\kilo\watt} and \SI{50}{\kilo\watt}. The core of the device is a force sensor, which consists of a commercial off-the-shelf EMFC weighing balance with \SI{10}{\micro\gram} readability and a mirror with high reflectivity ($R_L = $0.9998) attached to it. Another example of a recently developed device makes use of a miniature force sensor on the basis of flat-plate capacitive transducer technology using a single-reflection configuration, by which the best value obtained is approximately \SI{3.2}{\percent} expanded measurement uncertainty for the optical power levels of \SI{400}{\watt} (\SI{34}{\percent} at \SI{50}{\watt}) with \SI{260}{\milli\watt\per\sqrt{\hertz}} resolution \cite{Artusio_Glimpse_20}, although with \SI{1.6}{\percent} evaluated discrepancy between the photon-momentum measurement-based optical power calculations and reference optical measurements with thermopile detectors (as reported that has \SI{1.2}{\percent} measurement uncertainty in itself). A more recent development \cite{Artusio_Glimpse_2021}, having an integrated EMFC-balance based force sensor with similar \SI{10}{\micro\gram} readability, is already using an amplification effect by 14-reflection configuration. By this system a \SI{0.26}{\percent} and \SI{0.46}{\percent} relative expanded measurement uncertainty for power levels of \SI{10}{\kilo\watt} and \SI{1}{\kilo\watt} was reported, whereas using 1-reflection \SI{1.14}{\percent} and \SI{6.9}{\percent}, respectively.

In all above cases, the calibration of the force sensors is made against a discrete set of calibrated small mass artifacts ('transfer standard') ranging approximately as \SI{160}{\micro\gram}, \SI{420}{\micro\gram}, \SI{5}{\milli\gram}, \SI{6.7}{\milli\gram} up to \SI{30}{\milli\gram} and by further extraction of the calibration factor (or the correction factor) with the associated relative force sensor calibration uncertainty. The calibration of such mass artifacts is not a trivial task, requiring dedicated systems and specially controlled conditions, their handling and use as a transfer standard in calibrating other precision mass/force balances are troublesome. With one such system \cite{Shaw_2019}, that delivers high-precision SI-traceable mass/force calibration using concentric cylindrical capacitive transducers, the direct measurements of photon momentum generated forces have been performed as well. Using 1- and 7-reflection configuration with \SI{0.9}{\watt} input optical power a  direct force measurements at about \SI{17}{\nano\newton} and \SI{58}{\nano\newton} values was reported. Here, as relative combined expanded uncertainty of approximately \SI{5}{\percent} for 1-reflection and \SI{4}{\percent} for 7-reflection case, the systematic discrepancy of \SI{4}{\percent} as evaluated between the actually measured force values and the photon-momentum based force calculations from the referenced input optical power, using a power meter with \SI{5}{\percent} absolute error, due to cross-checking measurements against the radiometric transfer standards at \SI{1}{\watt}, are reported. Despite the reported values, here the consistence of the measurements given by the relative repeatabilities of half-nanonewton-level for the measurement of several ten nanonewton-level photon momentum forces from a \SI{1}{\watt} laser, demonstrate that for the macroscopic realization of mass at least at \SI{}{\micro\gram} to \SI{}{\milli\gram} levels the measurements can be complemented by a pure quantum mechanical effect 'photon-momentum' referenced by SI-traceable conventional power meters. Moreover, this would allow to achieve in perspective a fully SI-traceable true quantization of mass/force scale at \SI{}{\nano\gram}/\SI{}{\pico\newton}, \SI{}{\pico\gram}/\SI{}{\femto\newton} and below levels \cite{Kuck,Georgieva_Kuck_Lopez,Stephens_Lehman_2021,Williams_2019}. In other words, the realization of the mass/force scale through quantized energy, as quantified by the Planck constant.

\begin{figure*}
    \centering
    \includegraphics[width=0.8\linewidth]{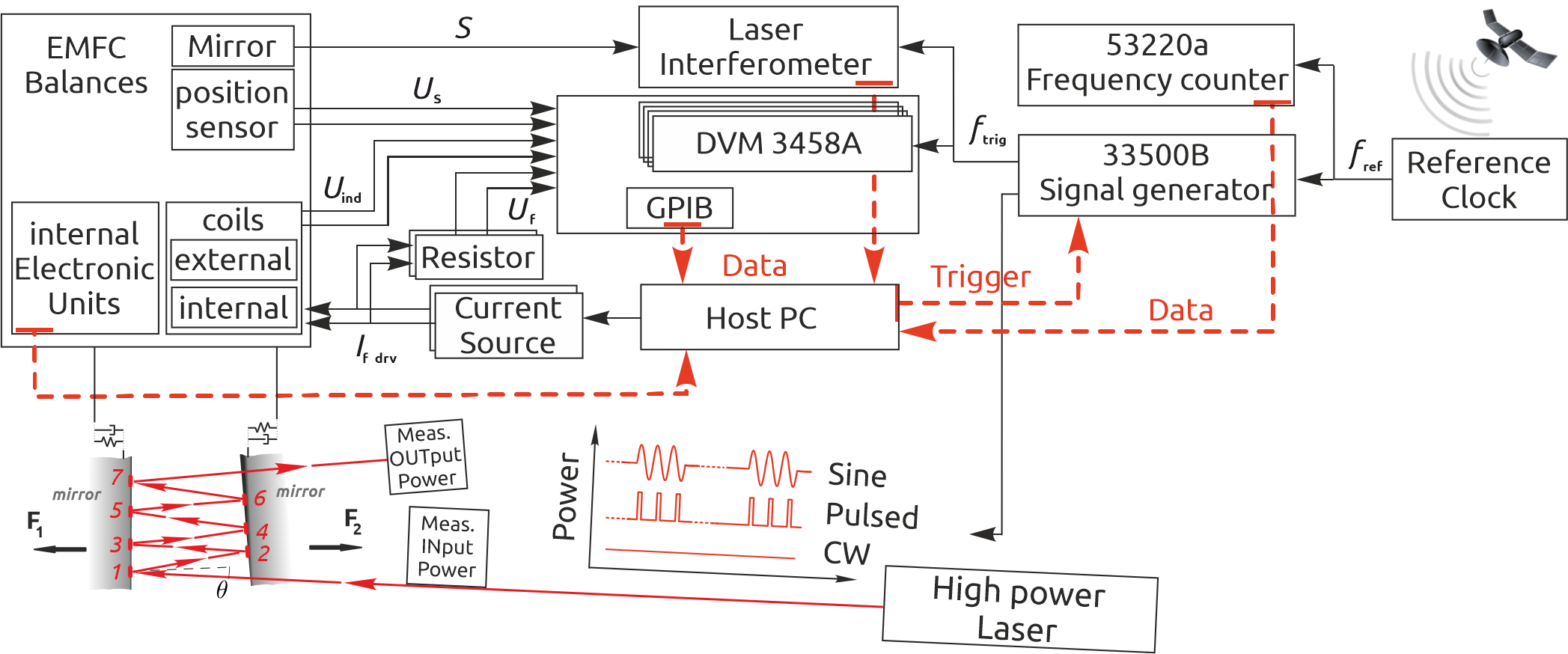}
    \caption{System blocks and signal flow diagram. GPS disciplined clock is used to reference the generated signal, which is used both to trigger synchronous voltage measurements (later also the length measurements) and to create the necessary type of function of modulated laser power. The precision current source and resistors are used to guide the “force mode” of the measurements.}
    \label{fig:fig6}
\end{figure*}

\begin{figure*}
    \centering
    \includegraphics[width=0.7\linewidth]{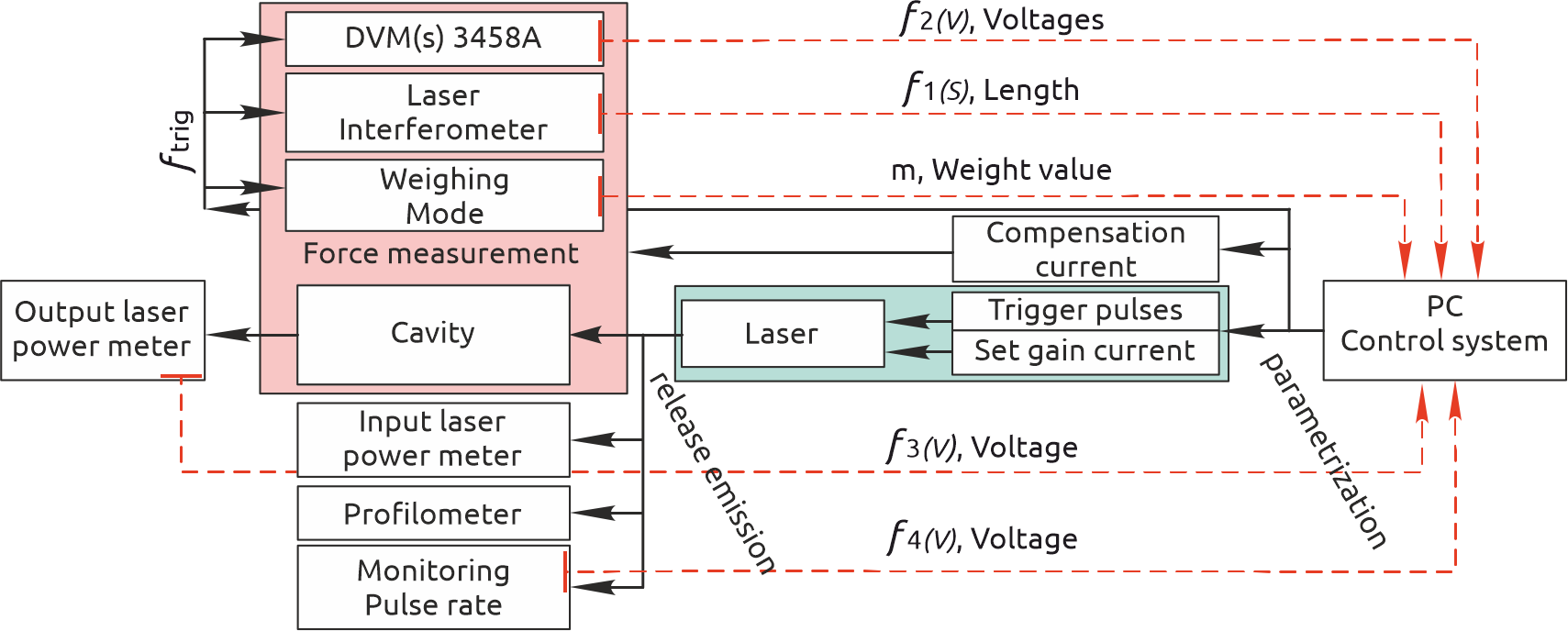}
    \caption{Overview of the main measurement quantities and the synchronization principle for measuring the voltages ($V$), length ($S$), and weight values ($m$) with their corresponding timestamp (frequencies - $f_1, f_2, f_3, ... $).}
    \label{fig:process_flow}
\end{figure*}

As expected, the measurement of optical power less than \SI{1}{\kilo\watt} using the measurements of photon momentum-generated forces in the case of a single reflection is very challenging because the force sensors need to resolve forces at least in the \SI{}{\nano\newton} range in an SI-traceable manner (note also that there is no broadly acceptable common force standard or widely used traceable reference-mass standards for this force ranges). To resolve this problem, irrespective of the absence of such a standard, we previously attempted \cite{OPEX_Vasilyan,Manske} to carry out force measurements and explore possibilities of reversed measurement routine by referencing them by SI-traceable optical power measurements, with no goal of making a comprehensive uncertainty budget assessment of the measurements. We used an early developed prototype with two force sensors (FMS) adapted for differential force measurements \cite{20nN}. With this FMS, the noise level was reduced by one order of magnitude from below \SI{1}{\micro\newton} level for a single sensor to below \SI{100}{\nano\newton} for a differential signal with \SI{20}{\nano\newton} resolution. Here, a multi-reflection configuration of a laser beam (with an optical power of approximately \SI{1}{\watt}) achieved in an optical cavity was demonstrated for the first time to generate calibration forces at the currently existing arguable lowest end  of the small force standard from \SI{10}{\nano\newton} up to \SI{10}{\micro\newton}, which are yet connected and are routinely being calibrated in relation to the mass standards, \SI{1}{\micro\gram} up to \SI{1}{\milli\gram}, respectively. Using a multi-reflection configuration, the total net force was amplified by at least one order of magnitude compared to the single-reflection configuration. In \cite{OPEX_Vasilyan}, forces up to \SI{150}{\nano\newton} with \num{21} reflections were achieved by a laser at an \num{800}-\SI{}{\milli\watt} power level. In \cite{Manske} already more deliberately made measurements confirmed the static and dynamic force measurement capabilities with the same FMS for up to \num{350}-\SI{}{\nano\newton} forces while using up to \SI{2}{\watt} laser power with maximum \num{22} reflection configuration. Here, \num{3} different lasers have been used with different wavelengths as sources for photon momentum force generation. As the reflective object suspended from FMS, a 1-inch square mirror based on fused silica substrate with a broadband dielectric coating providing approximately \SI{99.5}{\percent} reflectivity was used. Further, this portable FMS has been used to perform trial measurements at different metrology laboratories \cite{Vasilyan_2021} in collaboration with PTB (the laboratory for mass/force at the Technical University of Ilmenau and cleanroom laser radiometry laboratory at PTB, Division 4 Optics Department 4.5 Applied Radiometry). In addition to the earlier obtained results, for this extensive measurement campaign, a different laser source was used. A continuous-wave (CW) diode-pumped solid-state laser at a wavelength of \SI{532.50 \pm0.01}{\nano\meter} with a maximum output power of \SI{11}{\watt} (available at the PTB and also transported to TU Ilmenau for cross-comparison measurement purposes). The laser demonstrated itself with very good optical power stability of approximately \SI{0.5}{\percent} over \SI{2}{\hour}, although with a larger beam cross-section (Gaussian shape \SI{4}{\milli\meter} at $1/e^2$) compared to the previously used low-quality lasers. Several standard detectors (available at the PTB) for referencing the optical power of the laser beams that enter and leave the cavity have been used synchronously to monitor the possible power losses during the course of the entire force measurements. Here, the measurements were carried out for the three different multi-reflection configurations (\num{21}, \num{33}, and \num{41} reflections) for two different optical cavities, in each case with varying angles of incidence, angles of reflection, and patterns. The first cavity was created by using two conventional mirrors with $R_L$\,=\,\SI{99.5}{\percent} reflectivity, as previously, but of 2-inch size, which provided opportunities to increase the number of non-overlapping specular reflections. For the second cavity, a \num{2}-inch round ultra-high-reflectivity plane mirror was used with $R_L$\,=\,\SI{99.995}{\percent} reflectivity. In the latter, the reflective surface was optimized only for a wavelength of $\lambda$\,=\,\SI{532}{\nano\meter} by a multilayer dielectric coating on a synthetic fused silica substrate with a surface flatness below $\lambda / 10 $. These measurements led to a force measurement of approximately \SI{2000}{\nano\newton} at the upper limit, yielding approximately \SI{2.3}{\percent} of the overall relative standard uncertainty in the case of 33 reflections, also with reduced  systematic discrepancy of \SI{1.07}{\percent} between the actually measured force values and the photon-momentum based force calculations from the referenced input optical power. These results show a major improvement in the reliability of the method and the comparison of actual photon momentum-generated forces obtained from the real measurement data against those obtained by the idealized theoretical computations. The systematic error between the theoretically calculated forces and the forces obtained from the measurements, which was in the range of \num{10}-\SI{20}{\percent} in our earlier measurements \cite{OPEX_Vasilyan,Manske}, was reduced to less than \SI{1.7}{\percent} \cite{Vasilyan_2021} for the full range of obtained measurement data. The authors in [19] reported this difference to be approximately \SI{4}{\percent} for only two different input optical power values using an identical experimental configuration.

\begin{figure*}
    \centering
    \includegraphics[width=0.85\linewidth]{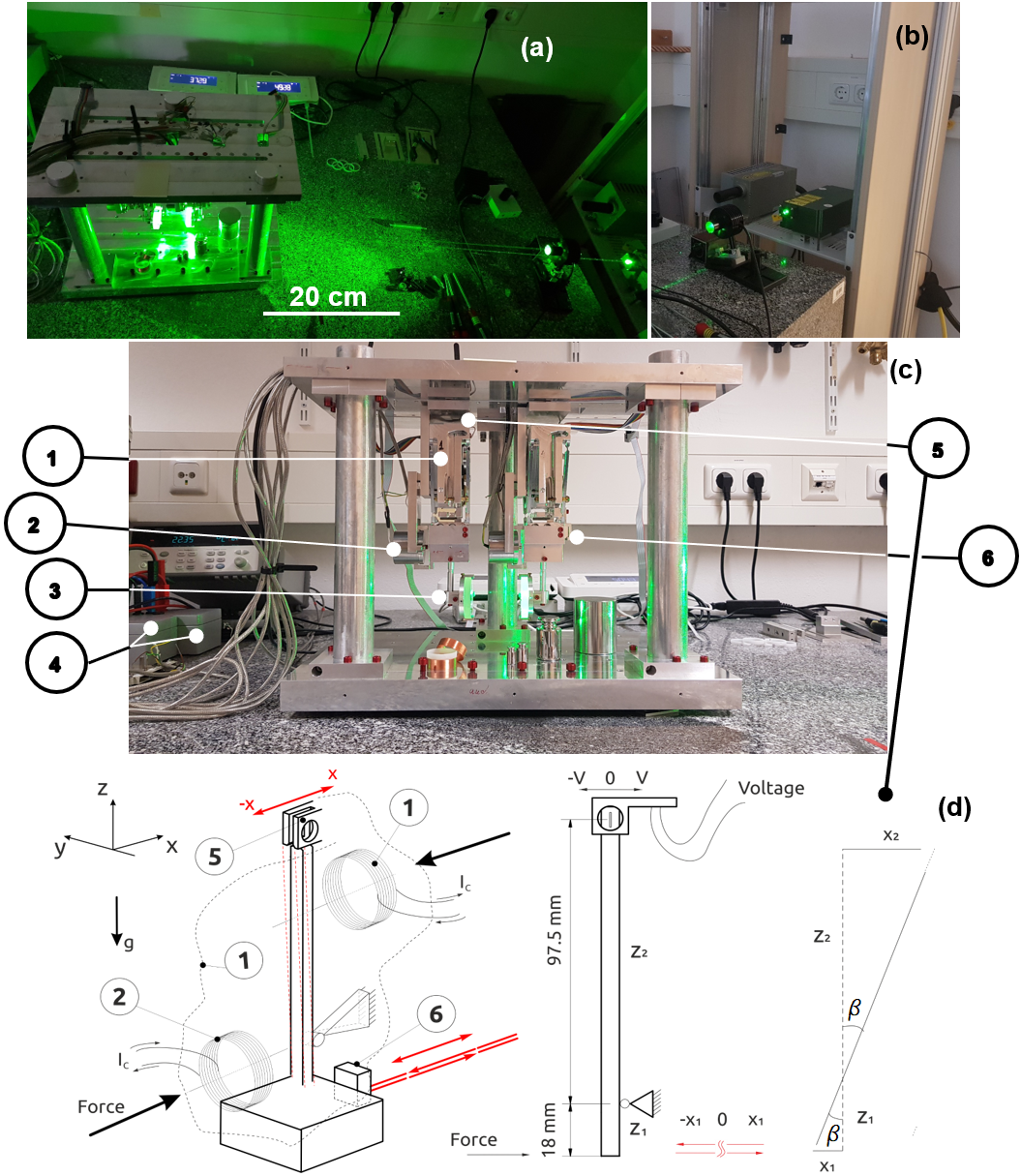}
    \caption{Photograph of the setup located at TU Ilmenau. \textit{a)} top view of the setup (assuming stable three contact points) placed on a massive granite block (metrology table) in the measurement laboratory, two grooves enable the EMFC balances to be positioned at different distances from each other. \textit{b)} Two pulsed diode lasers are used interchangeably to define the optical path for the traveling laser beam, creating a multi-reflection configuration inside the macroscopic cavity and eventually resting in the PTB-calibrated commercial (\textit{Ophire model: 10A-PPS}) optical power measurement sensor. \textit{c)} side view of the force-measurement system: \textbf{1} -- EMFC balance and position of the internally integrated magnet and coil assembly, \textbf{2} -- the external magnet and coil assembly, \textbf{3} -- mechanical adjustment system and the mirrors that form the macroscopic optical cavity, \textbf{4} -- precision resistors embedded in an isolation housing, \textbf{5} -- position sensor, \textbf{6} -- precision optical alignment half-cube fixed to the load carrier and used as interferometric measurement mirror. Purely for demonstration purposes for the size/scale cognition, in the foreground of the photograph at the bottom of the setup are shown (left to right) the baked varnish coils with a holder ring made of non-magnetic opaque white POM material and the standard weight pieces with \SI{10}{\gram}, \SI{20}{\gram}, \SI{200}{\gram}, and \SI{1}{\kilo\gram} nominal mass values. \textit{d)} Simplified schematics of deflection measurements based on the proportional lever arm structure, showing the calculation principle of the calibration/conversion factor for the position sensor in terms of \SI{}{\volt\per\meter}, (left-to-right) 3-D model, 2-D model (side view), and calculation principle scheme.}
    \label{fig:fig7}
\end{figure*}

\section{Design of the system and developed measurement infrastructure}
\label{design}

In this section, the details of the system design, the sub-components used in developments in an interchangeable manner based on a modular principle, and the developed measurement infrastructure are presented. In \cref{fig:fig6,fig:process_flow,fig:fig7}, the general operational schematics of the setup and its design are presented. One may notice several generalities adapted from the PB2 setup \cite{PB2_tm}. This FMS consists of two independently functioning EMFC balances (state-of-the-art precision weighing cells). They operate based on the principle of electromagnetic force compensation of the mechanically balanced proportional lever arm. The mechanical part of the balance is complete-monolithic, realized from a single piece of the material mechanism, with flexure hinges, adjustable or fixed counterweights, parallelogram guidance, and a transmission lever. The latter can be considered as a simple beam with a proportional lever arm structure that in the structure of the EMFC balance is supported by flexure hinges. The lever couples the mechanical system to the electromagnetic system, similar to coil actuators used in loudspeakers. Additionally, they are equipped with an optoelectronic absolute one degree-of-freedom (1-DOF) position-measuring sensor (which includes a light-emitting diode source, a light-receiver photo-detector, and a light-transmitting shutter slit), whose output electrical signal together with an analog or digital controller completes the guided measurement process in various operational regimes, including the open-loop, closed-loop, static, and dynamic regimes. The electromagnetic interaction is maintained by a highly optimized closed-circuit permanent-magnet system and a custom-manufactured coil actuator. To fit the best designing needs in obtaining the required electromagnetic $Bl$ factor in combination with semi-commercial closed-circuit magnet assembly, these coils are custom manufactured by a commercial vendor on the base of copper wire with various diameters ranging \SIrange{0.05}{0.5}{\milli\meter} as a baked varnish coil. This $Bl$ factor (also called “force factor”) in the actual Kibble balance experiment is determined through the force and velocity modes, \cref{eq:Fmg,eq:Uv}, and compared with each other at the level of several parts in \num{e-8} for mass values at \SI{1}{\kilo\gram} level \cite{Li_Schlamminger_irony}, unlike the table-top versions \cite{PB2_tm,chao2020performance} where the required precision in determination and consistency in comparison is at ppm level for example for \SI{1}{\gram} mass.

The supporting peripheral electronics were installed to record the internal position signal of the load cell and to supply electrical current to either of two coil actuators of each EMFC balance to be used interchangeably to maintain zero-point compensation in the “force mode” of operation by \cref{eq:Fmg} (“closed loop”, or “photon momentum mode”). In the “velocity mode” of operation (\cref{eq:Uv}), the induced voltage caused by an externally applied oscillatory motion is measured. The coil integrated inside the load cell, which is rigidly fixed to the proportional lever arm, we name \textit{internal coil}; the one which is rigidly fixed to the load carrier \textit{external coil}. Preliminary, several Agilent 3458A voltmeters \cite{Keysight3458A} were used to characterize and guide the measurements following the best metrological practices in precision voltage measurements. This, in combination with a programmable AC PJVS available at our laboratory in TU Ilmenau, obtains a direct SI-traceable calibrated voltage measurement for the 1-V range, with a measurement uncertainty typically below \num{3e-7}, which also provides the possibility to correct the absolute gain error and offset values. In addition, to obtain the value of the applied current (as an indication of the compensation force) the measurements of the voltage drop are arranged over a four-terminal standard-precision Vishay VHA518-11ZT resistor, we use interchangeably several resistors with different nominal values. Currently, the resistors integrated in the setup have nominal values of \SI{5}{\kilo\ohm} and \SI{10}{\kilo\ohm} and can be calibrated with an uncertainty below \num{1e-7}. During the course of the development, different external coils are used to achieve the necessary meticulous mechanical alignments with respect to the magnetic field of the magnet assembly. This is necessary also for the adjustments of the electrical measurements to the most desirable measurement ranges. The resistance value of these coils ranged from approximately \SI{3}{\kilo\ohm} and \SI{5}{\kilo\ohm} as a result of the material properties of the wire and its geometry. The number of turns is totaling in the range from \num{3000} to \num{5500}, in the volume comprising the coil by a height of \SI{10}{\milli\meter} and inner minimum and outer maximum diameters of \SI{29}{\milli\meter} and \SI{32}{\milli\meter}.

The magnet was selected in favor of a more conventionally used material with acceptable temperature stability - SmCo - that only has approximately one third the thermal coefficient of NdFeB; however, this choice was made at the expense of approximately \SI{20}{\percent} lower remanence than that of NdFeB. For a detailed description of the initial estimations and an overview of the initial development concept of the coil-and-magnet assembly see \cite{Schleichert_Vasilyan_magnet}.

In order to achieve complying requirements for the continuous measurement time, see \cref{fig:process_flow,fig:fig6}, for induced voltage, the voltage drop over the precision resistor, position sensor signal, interferometer (metrology laser) they are all triggered in a synchronous manner using a signal generator lock with an external reference GPS disciplined clock. In a similar manner, a synchronized bursts signal is employed to trigger the repetition rate $f_{rep}$ of the photon-momentum force generating pulsed laser modulations, having an additional control over the pulse width (temporal duration of the pulse $\tau$). The maintenance of this measurement process is be fully integrated in a single FPGA controller operation, the progress is underway. All of the measurement and the control of all sub-routines are yet accomplished by electronically programming separately each device and their interplay in a Matlab environment supported also by several nested sub-routines programmed in (C)C++.

{\renewcommand{\arraystretch}{1.2}
\begin{table*}
\centering
\caption{Parameters for tuning the settings of both lasers and for carrying photon momentum generated force measurements using all the modes. $^3$The repetition rate and pulse width were defined by different burst signals with varying values of duty cycle and supplied to the laser unit operating in an external control mode. The resulting parameters of the pulse train were controlled and measured in parallel when they reach the internal electronics of the laser units.}
    \begin{tabular}{l|c}
    \hline
         &  Measurement settings\\
            \hline
        Applied optical power in terms of CW type laser & Up to \SI{20}{\watt} (with \SI{<10}{\milli\watt} minimal increment) \\
        Types of optical power modulation  & Sine, square (duty cycle \SIrange{1}{99}{\percent}) $^3$ \\
        Modulation amplitude & \SI{0}{\watt} to min(\SI{2}{\watt}), (\SI{4}{\watt}), ..., max(\SI{20}{\watt}) (adaptable offset)\\
        Modulation frequency of gain current & \SIrange{0.0625}{2}{\hertz} (with \SI{0.02}{\hertz} increment) \\
        Number of periods & sine -- \num{5} - \num{15}, square -- \num{5} (gain modulated) \\
        Pulse modulation frequency & \SIrange{10}{625}{\kilo\hertz} \textbf{(IPG)}, \SIrange{10}{150}{\kilo\hertz}  \textbf{(VonJan)} \\
        Resolution of the pulse modulation frequency & \SI{2.5}{\kilo\hertz} \textbf{(IPG)}, \SI{0.5}{\kilo\hertz}  \textbf{(VonJan)}\\
        Pulse width & \SIrange{24}{100}{\nano\second} \textbf{(IPG)}, \SI{10}{\nano\second} to \SI{10}{\micro\second} \textbf{(VonJan)} \\
        Mirror reflectivity & \SI{99.995}{\percent} $\pm<$\SI{70}{ppm}\\
        Other optical transmission losses of cavity & $<$\SIrange{0.1}{0.5}{\percent}\\
        Number of reflections	& 21 and 33\\
         \hline
    \end{tabular}
    \label{tab:parametrization}
\end{table*}
}
The force measurement for a single EMFC balance defined for the open-loop mode are guided by the following measurement equation 
\begin{equation}
    F(t)=K_s\cdot S(t)=K_s\cdot k_U \cdot U_{pos}(t),
    \label{eq:open}
\end{equation}
and for the closed-loop
\begin{equation}
    F(t)=K_{Bl}\cdot I_{comp}=K_{Bl} \cdot \frac{U_f}{R} \text{, for}\,U_{pos}(t)=C,
    \label{eq:closed}
\end{equation}
where $K_s$ is the stiffness of the force measurement system, $S(t)$ is the displacement of the measurement point, $U_{pos}$ is the displacement measured as a voltage by the position sensor, $k_U$ is the sensitivity coefficient of the position sensor, $K_{Bl}$ is the electromagnetic force ($Bl$) factor, $I_{comp}$ is the compensation current necessary to apply to the coil to maintain the $U_{pos}$ at some original $C$ constant position. Assuming that two balances operate at the same measurement routine the differential signal can be constructed for the photon momentum generated force measurements in case of specular multi-reflections achieved in an optical cavity with both active sensing mirrors (see \cref{fig:fig6} lower panel) as
\begin{multline}
    F(t)=F_2(t)-F_1(t)=F_{P(Total)}= \\ (F_{P2}+F_{\varepsilon}) - (-F_{P1}+F_{\varepsilon})= \\ F_{P2}+F_{P1},
\end{multline}
where $F_{\varepsilon}$ is considered to be the common-mode noise in the signals of both EMFC balances, from which $F_{P1}$ and $F_{P2}$ photon momentum sensing forces can be also reconstructed by the force measurement equations given in \cref{eq:open,eq:closed}. The common-mode noise should be seen as a measurement noise that has two main components originating in one hand due to internal and external influencing factors on \text{optical-,} mechanical-, electrical- complex kinematics of the EMFC balances and on the other hand due to the complex light-matter interaction including the geometry of the laser-beam-reflections and real physical effects.
\begin{equation}
{F}_{\varepsilon}={F}_{\varepsilon}( \textbf{\footnotesize{\text{EMFC}}})+{F}_{\varepsilon}( \textbf{\footnotesize{\text{Laser}}}).
\label{eq:error}
\end{equation}
Based on \cref{eq:error} another configuration of photon-momentum force measurements can be constructed with only one photon-momentum force sensing EMFC balance while with the second one achieve a common-mode cancellation of at least the ${F}_{\varepsilon}( \textbf{\footnotesize{\text{EMFC}}})$ noise component, as
\begin{multline}
    F(t)=F_1(t)-F_2(t)=F_{P(Total)}= \\ ({F}_{\varepsilon}( \textbf{\footnotesize{\text{EMFC}}})) - (-F_{P1}+{F}_{\varepsilon}( \textbf{\footnotesize{\text{EMFC}}})+{F}_{\varepsilon}( \textbf{\footnotesize{\text{Laser}}}))= \\ F_{P1}-{F}_{\varepsilon}( \textbf{\footnotesize{\text{Laser}}}).
\end{multline}

Note that we do not account for two other possible noise components, mainly the cross-coupling between the EMFC balances and the optical stiffness of the cavity as a function of optical power and geometry of the mirrors and the laser beam.

In accordance with the simplified optical sub-schematics presented in \cref{fig:fig6} the input optical power values were used to theoretically calculate the expected photon momentum-generated forces, \cref{eq:2,eq:4}, for comparison with the data from the measurements. In this configuration, we can directly compare the reference for the force measurements and the reference for the optical power measurements towards the SI-traceable comparison of the force/mass measurements. Moreover, with such configuration the optical power detector (pre-calibrated by PTB with an overall expanded relative measurement uncertainty of \SI{<0.4}{\percent}) can be used to measure the optical power of outgoing laser beam and calculate the photon-momentum generated forces based on this residual optical power. The optical transmission losses as quantified previously (see figure 8 in \cite{Vasilyan_2021}) at about \SI{0.5}{\percent} have been reduced by deliberately aligning the laser beam reflections, at most of the multi-reflection configuration resulting to \SI{<0.1}{\percent}. This value, however, needs further multi-step verification because the optical power detector in itself has much larger measurement uncertainty. Note that this losses should theoretically be on the order of \num{500}\,ppm (i.e. \SI{0.05}{\percent}) in case of 10 laser beam reflections from the mirror with \SI{99.995}{\percent} reflectivity.

Two measurement campaigns were carried out using two different pulsed laser sources (in all previous measurements, only a CW laser was used) based on gain current regulated against the external or internal source diode-pumped solid-state technology at a wavelength of \SI{532}{\nano\meter}. The first laser (short: \textbf{VONJAN}) \cite{VonJan}, with a pulse-repetition rate ranging from \SIrange{10}{150}{\kilo\hertz} and beam diameter of \SI{4.2}{\milli\meter} and the pulse width optimized by an internal electronics to approximately \SI{17}{\nano\second} at \SI{30}{\kilo\hertz}, results in a maximum output average power of approximately \SI{9.2}{\watt}. The second (short: \textbf{IPG}) \cite{IPG}, with \SIrange{10}{625}{\kilo\hertz} repetition rate and beam diameter of \SI{2.6}{\milli\meter}, with approximately \SI{1.8}{\nano\second} pulse duration at \SI{600}{\kilo\hertz}, results in a maximum output average power of approximately \SI{20}{\watt} at \SI{625}{\kilo\hertz}. 

Initially, a special optical alignment strategy was adopted to achieve a defined number of specular reflections, \num{21} and \num{33}, after which the laser beam outgoing from the cavity was resting in the calibrated reference optical power meter. In the following we outline the main measurement routines for each of which the measurements using both lasers were performed: 
\begin{itemize}
    \item \textbf{Open loop} -— Measuring the effective deflection of the moving mirrors as a voltage obtained from the position sensor’s signal\footnote{ to be converted later to SI-traceable interferometric length measurement signal in order to obtain its sensitivity $k_U$, verify its stability, compensate for nonlinearities and provide the final position measurement uncertainty.} and the induced voltage in the coil (either internal or external) that is moving in the magnetic field of the permanent magnets. Within the open loop, the measurement can be made as 
    \begin{enumerate}
        \item a “\textbf{static}” measurement process, namely, measuring the deflection of the mirror and therefore the signal of the position sensor once the proportional lever of the EMFC balance is in its final rest at the steady-state. In this case, the induced voltage will be absent from the measurement routine. The photon momentum force will be defined by the measurement equations given in \cref{eq:2,eq:4}, Whereas the optical power will be defined through the \cref{eq:2,eq:4} based on actually measured force by $K_s \cdot S(t)$ product.
        \item a “\textbf{dynamic}” measurement process is used for defining the time-dependent function of the mirror’s motion (e.g., sine/square/pulsed function with a predefined number of periods, duty cycle, frequency, phase, and amplitude). This routine is commonly considered to be a “velocity mode” or “dynamic mode” in terms of the Kibble balance experiment \cref{eq:Uv,eq:10}. Here, the force that sets the system in an oscillatory motion can be chosen to be either the electromagnetic force acting by the second coil-magnet assembly or the photon momentum forces generated by the applied optical power.
    \end{enumerate}
    
    \item \textbf{Closed-loop force mode} —- Measuring the applied current to the coils (either external or internal) that is necessary to compensate the force acting on the loading carrier of the EMFC balance (or on the mirrors as a result of the photon momentum force) \cref{eq:Fmg}. In this case, there is no deflection-associated component in the measurement routine, and the measurements are solely guided by the voltage measurement resolution of the position sensor and the accuracy of the position control of the zero point with a digital or analog controller. For this case, the voltage measurement of the position sensor and the voltage drop over a precision resistor (current measurement) is necessary to create the input and output signals of the controller. This routine is commonly considered to be a “force mode” or “static mode” in terms of the Kibble balance experiment, when used in the vertical measurement direction to compensate for the gravitational force acting on the weight pieces.
    
    \item \textbf{Closed-loop weighing mode} —- Measuring (i.e., reading) the weight values from the factory electronics of the EMFC balances resulting from an implemented analog controller integrated in the electronics that again uses the position sensor and only the internal coil-and-magnet assembly \cref{eq:3}. Here, the weight values can be converted to a force value using the known value of the local gravitational acceleration at the site of the measurements, given that their scale is calibrated initially by the manufacturer against the standard weight pieces (only for the vertical measurement direction) using standard practice under high-class mass metrology laboratory conditions. Also, when the stiffness change of the system is compensated as well. Additionally, it is necessary to consider here the limiting factor that the measurement time is defined as the time duration after which the measured value oscillates only within a range of approximately 3 times the standard range of the static end weight value.
    
    \textbf{Closed-loop photon momentum mode} —- Measuring the optical power of the laser source with a calibrated reference/standard optical power measurement sensor, the beam of which is used to create multi-reflections in the cavity, then converting it to a force value using \cref{eq:2,eq:4}. The result can be used to determine the current necessary to compensate for the applied force acting on the loading carrier of the EMFC balance, \cref{eq:Fmg}. Note that the optical power can be measured as the average CW power for both types of CW and pulsed lasers, or in terms of the energy stored in a single pulse when either of average energy or average repetition rate of the laser pulse is defined, \cref{eq:15}. Additionally, one should assume all the possibilities of regulating the gain current, repetition rate, pulse width, and even accurately counting the photons flux rate (from the whole or fractional laser power) when considering the power in terms of the individual photons, \cref{eq:14}.
    
    \item \textbf{Closed-loop inverse photon momentum mode} —- a two-step mode \textit{i)} Measuring the applied current to the coils (either external or internal) \cref{eq:Fmg} that is necessary to compensate for the force acting on the loading carrier of the EMFC balance as a result of the photon momentum force, then \textit{ii)} using \textbf{Open loop} “\textbf{dynamic}” measurement process and obtaining the $Bl$ factor from the \cref{eq:10}. Here, the result can be converted to a force value using equation \cref{eq:11}, and therefore the optical power can be obtained using measurement equation given in \cref{eq:12}.
    
    \textbf{Closed-loop Reflectance measurement photon momentum mode} —- Measuring the applied current to the coils (either external or internal) \cref{eq:Fmg}, measuring the induced voltage and the oscillation velocity by \cref{eq:Uv,eq:10}, measuring the optical power of the laser source with a calibrated reference/standard optical power measurement sensor. And, combining all the measurement results in the measurement equation given in \cref{eq:16}. 
    
\end{itemize}

\begin{figure*}
    \centering
    \includegraphics[width=0.95\linewidth]{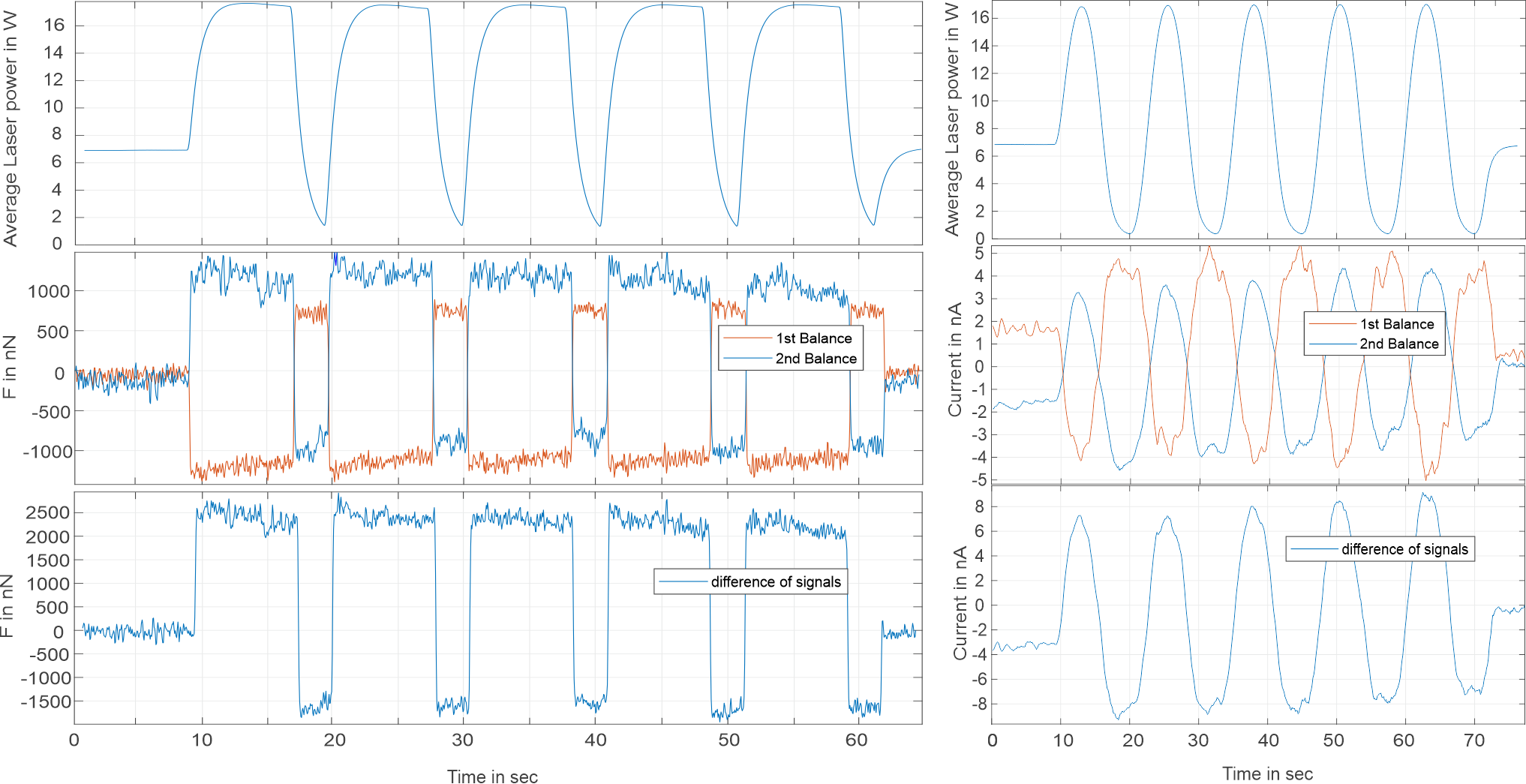}
    \caption{Closed-loop measurement using \textbf{IPG} laser and 33-reflection configuration. (\textit{left panel}) “weighing mode” measures the weight values from internal electronics of each EMFC balance then converting them to force values. In this example the optical power of the applied laser is modulated by changing the gain current in a square waveform with \num{0.1}-\SI{}{\hertz} frequency and \SI{75}{\percent} duty cycle using the function generator. (\textit{right panel}) “Closed-loop force mode” measures compensation currents for each EMFC balance applied to external coils which are used to maintain the compensation scheme (actually measured as a voltage drop over the \num{5}-\SI{}{\kilo\ohm} precision resistor, which is connected in series with a coil whose resistance is approximately \SI{5.03}{\kilo\ohm}). The value for the $Bl$ is roughly estimated to \SI{285.7}{\newton\per\ampere} (\SI{}{\tesla\meter}) using \cref{eq:Fmg}. In this example, the optical power of the applied laser is modulated by changing the gain current in sine waveform with frequency \SI{62.5}{\milli\hertz}. The power of the laser is measured as continuously averaged optical power (two graphs in the top panel) at the output of the cavity with a PTB calibrated reference thermopile sensor with $<$\SI{0.4}{\percent} ($k$=\num{2} coverage factor) expanded measurement uncertainty. The pulsed laser repetition rate was set at \SI{510}{\kilo\hertz}, which resulted in laser output power of approximately \SI{17}{\watt}. (\textit{top panel)} The optical transition losses of the cavity are measured to be typically $<$\SI{0.5}{\percent}. (\textit{middle panel}) The measurement signals obtained from both EMFC balance for each measurement configuration. (\textit{bottom panel}) Difference signals for each measurement configuration showing the measured force as a result of the total photon momentum transfer.}
    \label{fig:fig8}
\end{figure*}

\section{Measurement results}
In any of the measurement routines described in the previous section, the precisely determined $Bl$ factor (either external or internal coil-and-magnet assembly) underlies the possibility of direct SI-traceable calibration/determination of the force, the mass, or the optical power of the applied laser field using the photon momentum force generation method.

\begin{figure*}
    \centering
    \includegraphics[width=0.65\linewidth]{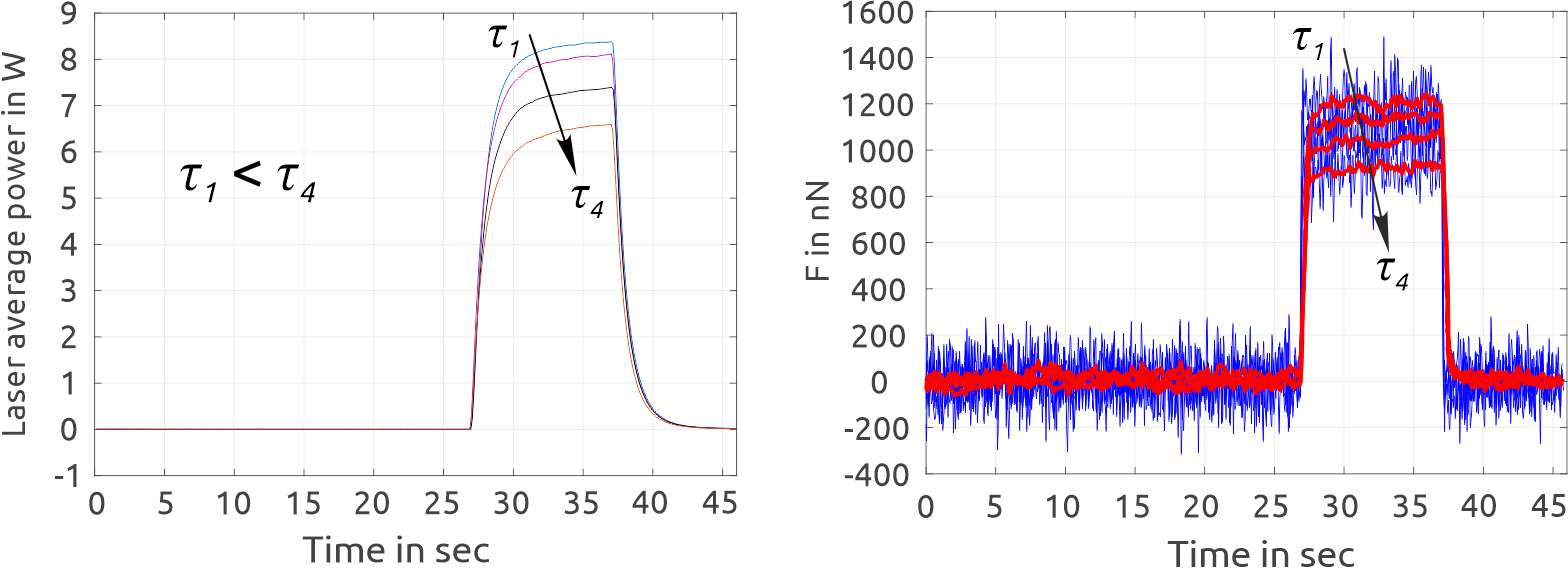}
    \caption{Closed-loop “weighing mode” measurement using \textbf{VonJan} laser and \num{21}-reflection configuration. In this example, the optical power of the applied laser is modulated by presetting the repetition rate with the signal generator at \SI{35}{\kilo\hertz} during the course of a \num{10}-\SI{}{\second} laser exposure time while varying only the pulse width $\tau$ by tuning the duty cycle to \SI{1}{\percent}, \SI{5}{\percent}, \SI{15}{\percent}, and \SI{25}{\percent} for each presented measurement as $\tau \approx$ \SI{286}{\nano\second}, \SI{1.43}{\micro\second}, \SI{4.29}{\micro\second}, and \SI{7.1}{\micro\second}. (\textit{left}) Average optical power measured for each trial by the thermopile sensor. (\textit{right}) Measured photon momentum-generated forces for different cases of the pulse width; only the measured difference signal for each measurement is presented as raw data (\textit{blue curves}) and filtered data (\textit{red curves})}
    \label{fig:fig9}
\end{figure*}

Figures (\ref{fig:fig8}) and (\ref{fig:fig9}), show only some of the typical sets of force and corresponding laser power measurements solely for the visual representation. The complete measurement configurations tested so far are given in parametric representation in \cref{tab:parametrization}. These measurements have been performed with a goal to verify the operational performance of the developed table-top system, identify the typical magnitudes of the measurement quantities and the possible errors arising due to chosen measurement strategies, and refine the developed measurement sub-routines. The maximum force values of approximately \SI{4000}{\nano\newton} with approximately \SI{32}{\nano\newton} resolution was measured at \SI{17}{\watt} input average optical power at 33-reflection configuration case. The simultaneous measurements of the same experimental process show obvious advantages of the photon-momentum method over the optical power measurement via thermopile detector, specifically in terms of the higher reaction time and the absence of the energy dissipation (heating) influences. An exemplary graph in \cref{fig:fig8} top-left demonstrates the underdeveloped measurement signal of the thermopile detector while the bottom-left is the same experimental process measured using the photon-momentum method. Some progress on the dynamic performance of the EMFC balances used for force (or optical power) measurements via the photon-momentum method is presented in our previous report \cite{Manske}. For the setup presented in the current work more detailed analysis will be provided in one of our future works.

\begin{figure}
    \centering
    \includegraphics[width=0.8\linewidth]{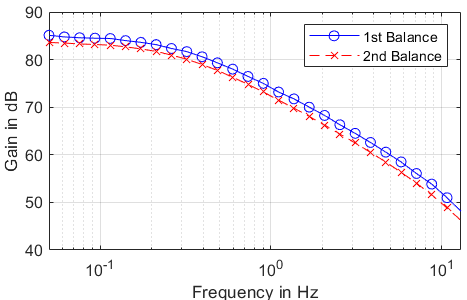}
    \caption{The transfer function of the position sensors integrated in the EMFC balances for different excitation frequencies. The input signal is the measured electric current (voltage drop over precision resistor) applied to each external coil in form of sine function with \SI{50}{}-\SI{}{\milli\ampere} amplitude and output signal is the measured voltage change of the position sensors from each EMFC balance.}
    \label{fig:pos_sensor}
\end{figure}

Additional experimental tests have been performed for the length measurements, for which only the uncalibrated optical position sensor was used so far. These position sensors are custom adjusted for each EMFC balance separately; therefore, a separate calibration should be made with an interferometric length measurement system that would additionally serve to create SI traceability. Initially, it is necessary to quantify the precision and the absolute amplitude to which the laser interferometer needs to be adjusted and later resolve the deflection measurements. In \cref{fig:fig8} it is demonstrated the typical range of measurements, particularly for the force levels up to \SI{10}{\micro\newton} the necessary compensation currents to be applied to the external coils ranges in the order of \SI{35}{\nano\ampere}, the position sensor voltage measurements about \SI{1}{\milli\volt}, and thus, for the actual deflection measurements should be considered the range of \SIrange{0.1}{2}{\micro\meter}, see \cref{fig:fig7}d. In \cref{fig:pos_sensor}, we present the current results of the experimentally measured transfer function of both position sensors against an externally applied force for different frequency components.

\begin{figure}
    \centering
    \includegraphics[width=0.75\linewidth]{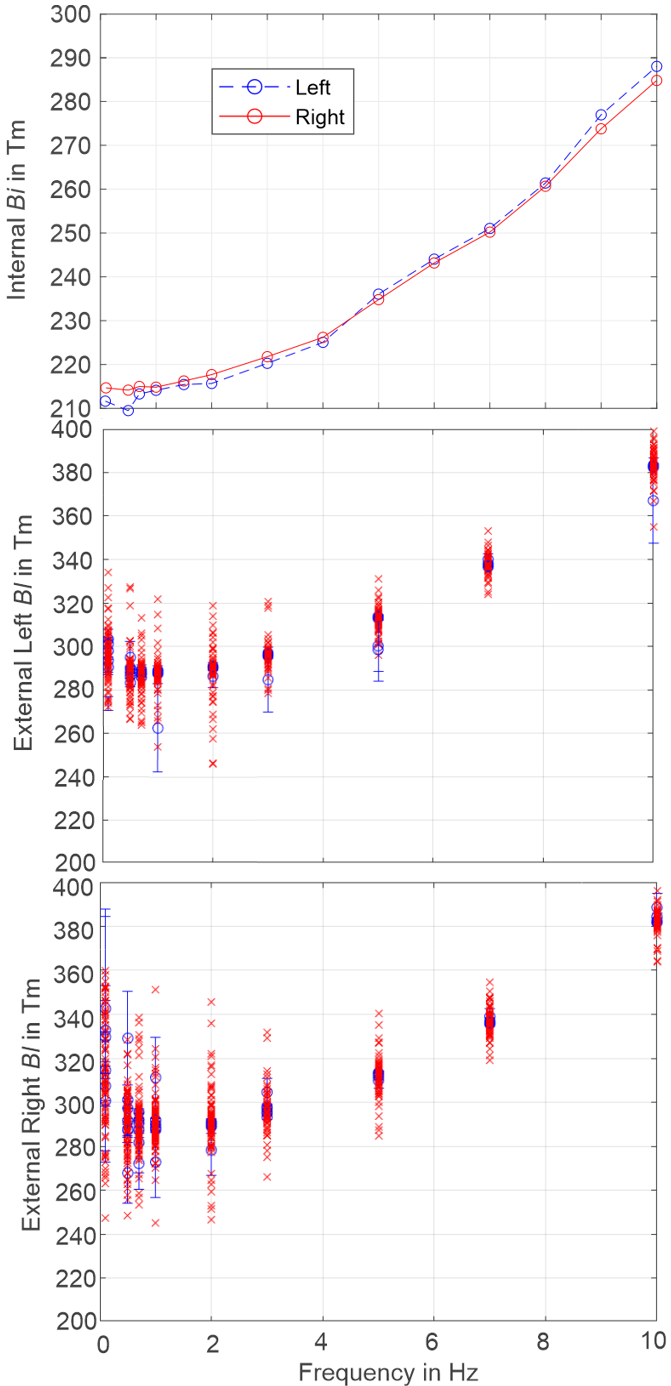}
    \caption{$Bl$ factors of coil–magnet assemblies for different excitation frequency components. (\textit{top}) Internal coil–magnet assembly: the oscillation was created with approximately \SI{10}{\micro\ampere} electric current applied to an external coil–magnet assembly. (\textit{middle and bottom}) External coil–magnet assemblies attached from each EMFC balance: the oscillations was created with each internal coil–magnet assembly of EMFC balances with varying amplitude of \SI{10}{\micro\ampere}, \SI{25}{\micro\ampere}, \SI{50}{\micro\ampere}, \SI{100}{\micro\ampere}, \SI{150}{\micro\ampere} applied electric current (raw values - '$\times$', mean and standard deviation for each amplitude '$\circ$---'). An electric current divider was used to reduce the initially applied relatively high electric currents.}
    \label{fig:Bls}
\end{figure}

Based on the developed electrical measurement infrastructure some systematically conducted measurement campaigns for the force mode and velocity mode measurement routines described in previous sections are made. The results of these investigations provide a basis for the development of evaluation strategies for the raw measurement data obtained by the interferometer, voltmeters, and optical power meter for the final identification of the $Bl$ factors of internal and external coil–magnet assembly. This includes systematic improvements related to the data acquisition options, carrying out more measurements at different frequencies for the velocity mode typically at \SIrange{0.05}{10}{\hertz} to refine the measurement uncertainties and the non-linearity corrections, identification of parasitic torque and tilt effects whose influences are coupled in both, in the oscillatory motion and during the static measurement conditions. Additionally, the system is adjusted to allow the identification of various effects, related to the coil immersion depth- and lateral position-dependent variations in the magnet system, apparent in measurements of the $Bl$ factors. In accordance with the data flow schematics provided in \cref{fig:fig6,fig:process_flow} and preliminary chosen data evaluation strategies from \cref{eq:Fmg,eq:Uv,eq:7,eq:8,eq:9,eq:10} we obtained the $Bl$ factors of all (internal and external) coil–magnet assemblies. In \cref{fig:Bls}, we present the results of the measurements, which are normalized approximately for a \SI{1}{}-\SI{}{\hertz} frequency and based on our pre-knowledge in a typical value of the sensitivity factor of each position sensor (the ratio of the position-voltage to absolute length change values). Despite the reproducible and stable performance of the measurement system and the measurement routine, the results are yet to be calibrated using interferometric length measurement. Here, the non-linearities of the position sensor and the characteristic transfer functions of both the lever arm (mechanical) and of the coil-and-magnet assemblies (electromagnetic) are apparent and need to be corrected. Not that an elaborated high-precision multi-harmonic sine fitting algorithm has to be applied in order to reduce and correct for the non-linear effects as well as the numerical errors. The bases for the sine fitting algorithm used for the currently obtained measurement results serve \cref{eq:7,eq:8,eq:9,eq:10} as the case of ideal single-component sine waves which represent the simplified form, however, a more complete form and the exact implementation principle is provided in our previous publication \cite{PB2_tm} that is in progress to be adapted for this system as well.

\section{Conclusion}
\textbf In this report we have presented the design and operational performance of table-top measurement apparatus combining Kibble balance principle and photon momentum method with multi-reflected laser beam configuration towards direct Planck constant traceable high accuracy and high precision small forces and optical power measurements within SI unit system. This hybrid quantum-opto-electro- and classical mechanical system, together with the photon momentum generated force measurement method, should primarily serve to obtain a unifying scaling system in the fields of high optical power and force metrology based on one of the most fundamental quantity, that is, the force via momentum-energy conservation law. Purely optically generated photon momentum forces of up to approximately \SI{4000}{\nano\newton} values at input average optical power of laser approximately \SI{17}{\watt} and 33-reflection configuration was achieved. The configurations of this specially tailored optical, mechanical, electrical system was adapted to operate at the limiting cases where the photon momentum method would serve to directly connect in continues scale the macroscopic realization of force, weight and optical power measurements with the microscopic quantum physical effect. In our previous publication \cite{Vasilyan_2021}, we presented the general scaling and the estimation of the expected measurement uncertainties for this measurement method which have guided us in this work, and together with the advance presented in current work will be the base for the forthcoming complete measurement uncertainty estimations. Additionally, a comprehensive overview of the theoretical background, the notable achievements in this field and the practically realizable state-of-the-art metrological framework are detailed towards truly quantized SI traceable forces measurements by photon momentum method.

\section{Acknowledgment} This work has been funded in part by DFG, and EUROSTARS-2 joint programme co-funded by EU research and innovation programme. We are grateful to a number of industrial partners and collaborators for providing technical support. Sartorius AG for providing the EMFC balances, Supracon AG for the primary AC and DC programmable Josephson voltage standard system, PTB for the calibration of some of the precision resistors, and the optical power measurement sensor. IPG Laser GmbH and VONJAN Technology GmbH for providing the opportunity in operating with the pulsed lasers and sharing with the details of the original factory developed internal electronics. The authors acknowledge all the constructive discussions made with the colleagues E. Manske and F. Hilbrunner from the Institute of Process Measurement and Sensor Technology at TU Ilmenau.

\bibliographystyle{unsrt}
\bibliography{Pulsed_laser_Vasilyan_references}


 




\end{document}